\documentclass[]{pasj01}

\usepackage{color}

\begin{document} 
\Received{}
\Accepted{}

\title{X-Ray Bright Optically Faint Active Galactic Nuclei in the Subaru Hyper Suprime-Cam Wide Survey}


\author{Yuichi \textsc{Terashima} \altaffilmark{1}%
}
\altaffiltext{1}{Department of Physics, Ehime University, Bunkyo-cho, Matsuyama, Ehime 790-8577, Japan}
\email{terasima@astro.phys.sci.ehime-u.ac.jp}
\author{Makoto \textsc{Suganuma}\altaffilmark{1}}
\author{Masayuki \textsc{Akiyama},\altaffilmark{2}}
\altaffiltext{2}{Astronomical Institute, Tohoku University, Aramaki, Aoba-ku, Sendai, Miyagi 980-8578, Japan}
\author{Jenny E. \textsc{Greene}\altaffilmark{3}}
\altaffiltext{3}{Department of Astrophysics, Princeton University, Princeton, NJ 08544, USA}

\author{Toshihiro \textsc{Kawaguchi}\altaffilmark{4}}
\altaffiltext{4}{Department of Economics, Management and Information Science, 
Onomichi City University, Onomichi, Hiroshima 722-8506, Japan }
\author{Kazushi \textsc{Iwasawa}\altaffilmark{5,6}}
\altaffiltext{5}{Institut de Ci\`encies del Cosmos (ICCUB), Universitat de Barcelona (IEEC-UB), Mart\'i i Franqu\`es, 1, 08028 Barcelona, Spain}
\altaffiltext{6}{ICREA, Pg. Llu\'is Companys 23, 08010 Barcelona, Spain}
\author{Tohru \textsc{Nagao}\altaffilmark{7}}
\altaffiltext{7}{Research Center for Space and Cosmic Evolution, Ehime University, Bunkyo-cho, Matsuyama, Ehime 790-8577, Japan}
\author{Hirofumi \textsc{Noda}\altaffilmark{8}}
\altaffiltext{8}{Frontier Research Institute for Interdisciplinary Sciences, Tohoku University, Aramaki, Aoba-ku, Sendai, Miyagi 980-8578, Japan}
\author{Yoshiki \textsc{Toba}\altaffilmark{9}}
\altaffiltext{9}{Academia Sinica Institute for Astronomy and Astrophysics, P.O. Box 23-141, Taipei 10617, Taiwan}
\author{Yoshihiro \textsc{Ueda}\altaffilmark{10}}
\altaffiltext{10}{Department of Astronomy, Graduate School of Science, Kyoto University, Kitashirakawa-Oiwake cho, Sakyo-ku, Kyoto, Kyoto 606-8502, Japan}
\author{Takuji \textsc{Yamashita}\altaffilmark{7}}


\maketitle

\begin{abstract}
We construct a sample of X-ray bright optically faint active galactic nuclei by combining Subaru Hyper Suprime-Cam, {\it XMM-Newton}, and infrared source catalogs. 53 X-ray sources satisfying $i$ band magnitude fainter than 23.5 mag and X-ray counts with EPIC-PN detector larger than 70 are selected from 9.1 deg$^2$, and their spectral energy distributions (SEDs) and X-ray spectra are analyzed.44 objects with an X-ray to $i$-band flux ratio $F_{\rm X}/F_i>10$ are classified as extreme X-ray-to-optical flux sources. SEDs of 48 among 53 are represented by templates of type 2 AGNs or starforming galaxies and show signature of stellar emission from host galaxies in the optical in the source rest frame. Infrared/optical SEDs indicate significant contribution of emission from dust to infrared fluxes and that the central AGN is dust obscured. 
Photometric redshifts determined from the SEDs are in the range of 0.6--2.5. X-ray spectra are fitted by an absorbed power law model, and the intrinsic absorption column densities are modest (best-fit $\log N_{\rm H} = 20.5-23.5$ cm$^{-2}$ in most cases). The absorption corrected X-ray luminosities are in the range of $6\times10^{42} - 2\times10^{45}$ erg s$^{-1}$. 20 objects are classified as type 2 quasars based on X-ray luminsosity and $N_{\rm H}$. The optical faintness is explained by a combination of redshifts (mostly $z>1.0$), strong dust extinction, and in part a large ratio of dust/gas. 

\end{abstract}

\section{Introduction}

Supermassive black holes (SMBHs) lurking in the nuclei of galaxies have grown through mass accretion and merging. 
SMBHs are observed as active galactic nuclei (AGNs) and the peak of their number densities
are at redshift $z\sim2$ for quasars and 0.7--0.8 for Seyferts (Ueda et al. 2003, 2014, Hasinger et al. 2005). 
The redshifts around these peaks are a key epoch in understanding the evolution of
SMBHs. In order to sample whole populations of SMBHs in this epoch, 
it is mandatory to perform multi-wavelength surveys to avoid selection biases as much as possible. 
In particular, obscured AGNs, which are believed to constitute a large fraction of AGNs and a key stage of AGN evolution, 
are biased against conventional surveys using the UV and optical bands.

Modern surveys combining the optical, infrared, and X-rays indeed reveled the presence of various classes of AGN that could be missed in conventional optical surveys.
Extragalactic X-ray surveys have shown the presence of optically faint populations 
with X-ray (2--10 keV) to optical ($R$ band) ratio $F_{\rm X}/F_R>10$, 
which are more than one order of magnitude fainter in the optical compared to ordinary AGN populations (e.g., Brandt \& Hasinger 2005).
Fiore et al. (2003) pointed out that 20\% of X-ray sources found in {\it Chandra} or {\it XMM-Newton} surveys are $F_{\rm X}/F_R>10$.
Such an X-ray bright but optically faint population is of great importance in understanding the nature and evolution of AGNs. 
The optical faintness could be caused by significant dust extinction, 4000 {\AA}/Balmer break or Lyman break 
shifted to infrared bands, or a combination of both (e.g., Hornschemeier et al. 2001, Rigby et al. 2005).
The population with a significant amount of dust might be AGNs with the ongoing growth embedded in a large amount of gas and dust
(e.g., Hopkins et al. 2008). 
If 4000 {\AA}/Balmer break is shifted to 
infrared band, redshifts are inferred to be greater than unity that coincides with 
the peak of the number density of luminous Seyfert galaxies and quasars. 

X-ray bright optically faint AGNs are found and studied in various X-ray surveys.
Brand et al. (2006) studied X-ray and optical properties of 773 {\it Chandra} X-ray sources with 
$F_{\rm X}/F_R>10$ in the XBo\"otes field of 9.3 deg$^2$ and 
found that they have redder color in the optical and harder X-ray spectra than 
those of 
the X-ray source population with $0.1<F_{\rm X}/F_R<10$, 
where X-ray flux is measured in the 0.5--7 keV band.
This result implies that the optically faint sources are obscured AGNs, 
although detailed studies of individual X-ray spectra are not possible for most objects because of the shallow {\it Chandra} pointings of 5 ksec exposure.

Rigby et al. (2005) compiled 20 X-ray sources with $F_{\rm X}/F_R>10$ in  a part of the {\it Chandra} Deep Field South 
and show that most of them (17/20) indicate an apparently flat X-ray slope ($\Gamma < 1.4$).
Civano et al. (2005) present results of analysis of stacked X-ray spectra of high X-ray/optical ratio sources 
in the {\it Chandra} Deep Fields North and South and found a very flat slope with a photon index of $\approx$ 1.0 independent of source flux. 
Individual spectral fits of brightest objects imply absorption column densities of $10^{22} - 10^{23.5}$ cm$^{-2}$. 
Rovilos et al. (2010) analyzed X-ray spectra of optically faint sources and found  that majority (27/35) are absorbed by $N_{\rm H} >10^{22}$ cm$^{-2}$. Three among them are candidates for Compton-thick AGNs. The stacked spectrum of the 35 sources is flat with a photon index of $\approx 0.9$.

Perola et al. (2004) analyzed X-ray spectra of 24 objects with $F_{\rm X}/F_R>10$ found in the HELLAS2XMM survey 
covering 1 deg$^2$ with a flux limit of $F_{\rm X}\approx10^{-14}$ erg s$^{-1}$ cm$^{-2}$ in 2--10 keV. 17 among them 
are absorbed by a hydrogen column density of $N_{\rm H} = 10^{22} - 6\times10^{23}$ cm$^{-2}$, 
while  $N_{\rm H}$ for the rest are less than $10^{22}$ cm$^{-2}$. 
They found also that the fraction of highly absorbed sources becomes greater for larger $F_{\rm X}/F_R$ sources. 
Tajer et al. (2007) found that 20\% of 124 X-ray sources with measured $N_{\rm H}$ in the {\it XMM-Newton} Medium Deep Survey (XMDS),
which covers 1 deg$^2$ with a flux limit of $F_{\rm X}\approx10^{-14}$ erg s$^{-1}$ cm$^{-2}$, have $F_{\rm X}/F_R>10$. 21 among 25 objects with 
$F_{\rm X}/F_R>10$ are classified as type 2 quasars ($N_{\rm H}>10^{22}$ cm$^{-2}$ and X-ray luminosity 
in 2--10 keV $L_{\rm X}>10^{44}$ erg s$^{-1}$). 

Della Ceca et al. (2015) utilized the second {\it XMM-Newton} serendipitous source catalogue and optical imaging data 
to find seven extreme optical-to-X-ray ratio objects
($F_{\rm X}/F_R>50$)  with an X-ray flux greater than $1.5\times10^{-13}$ erg cm$^{-2}$ s$^{-1}$. 
Three and two of the seven sources are classified as type 2 quasar, and BL Lac 
objects, respectively. The rest are unidentified but indicate obscured  quasar nature. 
The X-ray spectra of the three identified type 2 quasars are represented by a power law absorbed by Compton-thin matter.
Thus previous observations of optically faint AGNs suggest that most of them are absorbed AGN, while less absorbed AGNs are also present.

Optical to infrared spectral energy distributions of X-ray bright optically faint sources tend to be very red (Rigby et al. 2005).
$R-K$ or $V_{606}-K_s$ colors of majority of them are extremely red ($>5$) (Rovilos et al. 2010, Brusa et al. 2010). 
Optically faint sources also tend to be bright in mid-infrared relative to the optical bands if detected in the mid-infrared band.
24 $\mu$m to $R$-band flux ratios ($F_{24}/F_R$) of some sources exceed 1000 
(Lanzuisi et al. 2009, Brusa et al. 2010, Della Ceca et al. 2015), 
which is a criterion of dust-obscured galaxies (Dey et al. 2008). 
Among 43 objects with $F_{24}/F_R>2000$ in the sample of Lanzuisi et al. (2009), 
at least 23 are X-ray bright optically faint for example.
Thus strong infrared emission relative to the optical is also a property of this population
and implies an important role of dust obscuration in at least objects with infrared detections.

In this paper, we report a new sample of X-ray bright optically faint AGNs selected by combining {\it Subaru} Hyper Suprime-Cam (HSC: Miyazaki et al. 2012) Subaru Strategic Program (SSP: Aihara et al. 2017a, b), 
{\it XMM-Newton},  and infrared surveys ({\it Spitzer} and UKIRT Infrared Deep Survey), 
and  discuss their nature.  The large area and survey depth of HSC enable us to 
construct a large sample of rare populations and to constrain the spectral energy distributions (SED) of optically faint sources.
This paper is organized as follows. Section 2 describes the data sets. Section 3 explains the procedure of the sample selection.
Sections 4 and 5 provide results of analysis of SEDs in the optical and infrared bands, and X-ray spectra, respectively.
Discussions about the nature of the sample is given in Section 6. Section 7 summarizes our findings.
We adopt cosmological parameters of $H_0=70$ km s$^{-1}$ Mpc$^{-1}$, $\Omega_{\rm M}=0.3$, and
$\Omega_\Lambda=0.7$. All magnitudes refer to the AB system.

\section{The data}

\subsection{X-ray}

The {\it XMM-Newton} Large Scale Structure Survey (LSS: Pierre et al. 2004) and its extension 
(the Ultimate {\it XMM-Newton} survey, XXL; Pierre et al. 2016) 
are X-ray surveys covering a large consecutive fields (25 deg$^2$ in the northern field), which enable us to study large scale structures and to find rare X-ray emitting populations.
We use a part of the XXL northern field covered by HSC and {\it Spitzer} space telescope.
The catalogs used in our analysis are summarized in table 1.
The XMM serendipitous source catalogue data release 6 (3XMM-DR6: Rosen et al. 2016) is used
as an X-ray source catalog. 
The 3XMM-DR6 catalog utilizes {\it XMM-Newton} observations performed by 2015 June 4, and contains some pointed observations in the XXL region with a longer 
exposure time than those of the original survey. Thus we fully utilize X-ray sources detected in such additional pointings.
Since our aims are to select 
X-ray bright optically faint sources and understand their nature, 
we use objects with X-ray counts 
obtained with the EPIC-PN detector in 0.2--12 keV listed in the 3XMM-DR6 catalogue 
greater than 70 so that X-ray spectral fits can be performed. This count limit roughly corresponds to an X-ray flux of $\sim10^{-14}$ erg s$^{-1}$ cm$^{-2}$ in the 2--10 keV band, 
though conversion to an X-ray flux depends on exposure time and assumed X-ray spectra.
If an X-ray source is observed more than twice, the observation giving largest X-ray counts is used.

We retrieved Observation Data Files (ODF) of {\it XMM-Newton}  for all the candidates for optically faint population selected in section 3. 
The ODF were reprocessed using the {\it XMM-Newton} Science Analysis Software (SAS) version 14.0.0 and 
Current Calibration Files (CCF) as of 2015 Oct. 20. 
We made light curves of whole field of view in the 10--12 keV excluding bright X-ray sources to examine 
stability of background, where only PATTERN 0 events were used. 
We excluded time intervals with high background rates. 
Then we made X-ray images in the 0.5--2, 2--10, and 0.5--10 keV bands, and examined the brightness of candidate X-ray sources
selected by matching X-ray, 3.6 $\mu$m, and $i$ bands as described in section 3.3.
 In some cases, no X-ray source is clearly visible in the images and are excluded from our sample. 
 If the X-ray source counts after the data screening
 become smaller than 70 counts per EPIC-PN, they are also excluded.

\subsection{Subaru HSC}

We use the optical imaging survey performed with HSC.
The XMM-XXL region is covered by the wide layer of the HSC survey in the ongoing SSP.
The wide layer of the HSC survey conducts imaging with the five bands ($g, r, i, z,$ and $y$), for which the limiting magnitudes (5$\sigma$) are 
26.8, 26.4, 26.4, 25.5, and 24.7, respectively. The source catalogs selected in the $i$ band and images taken from 
the data release S15B are used in the following analysis. 
All the magnitudes are corrected for Galactic extinction according to Schlegel et al. (1998).
Cmodel magnitudes are used in all the analysis unless otherwise noted.

Among the five filter bands, we use $i$-band data for our sample selection because of the following reasons. 
Most of previous studies of X-ray bright optically faint populations utilized the $R$ band. By using the $i$ band centered at a longer wavelength,
we expect to detect objects with a higher redshift if the cause of the optical faintness is the 4000 \AA/Balmer break or Lyman break.
Secondly, the limiting magnitude is the best among the three bands ($i, z$, and $y$) redder than the $R$ band. 
Thirdly, the $i$-band data are 
taken under a good seeing condition with a point source FWHM $<$
\timeform{0".7}, which enables us to better quantify extendedness.
After performing sample selection using $i$ band, all the five band data are used to study SEDs and photometric redshifts.

\subsection{Infrared}

Infrared data are useful to better identify more likely counterparts of X-ray sources as used in multi-wavelength surveys 
(e.g., Brusa et al. 2007, Civano et al. 2012, Akiyama et al. 2015, Marchesi et al. 2016).
A part of the XMM-XXL region is observed by the {\it Spitzer} Infrared Array Camera (IRAC) and Multiband Imaging Photometer (MIPS).
We required X-ray and HSC sources are in the area covered by the {\it Spitzer} IRAC 3.6 $\mu$m band public catalog of
the {\it Spitzer} Wide-Area Infrared Extragalactic Survey (SWIRE: Lonsdale et al. 2003, 2004, Surace et al. 2005). 
This requirement determines the area used in this study
(9.1 deg$^2$). This sky region is fully covered by 3XMM-DR6 and HSC, and  partly overlaps with
the Subaru/XMM-Newton Deep Survey (SXDS: Akiyama et al. 2015, Ueda et al. 2008)
or XMM-Newton Medium Deep Survey (XMDS: Chiappetti et al. 2005, Tajer et al. 2007).
Nearly same region is also covered by the 4.5, 5.8, 8.0 $\mu$m IRAC bands and 24 $\mu$m MIPS band. 
These multiband data are also used if available.
We used aperture photometry with a radius of 1.9" after aperture corrections.
The limiting flux ($5\sigma$) for the five bands are 
7.3, 9.7, 27.5, 32.5, and 450 $\mu$Jy, respectively.

Near infrared photometric data obtained in the UKIRT Infrared Deep Sky Survey (UKIDSS: Lawrence et al. 2007)
are also used to better constrain the spectral energy distributions and photometric redshifts, if available. 
The photometric catalog of the UKIDSS data release 10 plus for the two survey layers,
Deep Extragalactic Survey (DXS) and Ultra Deep Survey (UDS), are used in our analysis.
Photometric data in the $J$ and $K$ bands
are available for the former and the latter is covered by the $J, H,$ and $K$ bands.
The limiting magnitudes ($5\sigma$) of DXS are 
23.2--23.3 ($J$ band) and 21.2--22.8 ($K$ band) (Warren et al. 2007),
while those of UDS are 25.7, 25.2, and 24.8 for the $J$, $H$, and $K$, bands, respectively.
All the UKIDSS photometric data are corrected for Galactic extinction (Schlegel et al. 1998).

\section{Selection}

\subsection{Candidates for X-ray Bright Optically Faint Sources}

Candidates for X-ray bright optically faint sources were selected by matching X-ray and $i$-band sources. 
In the following procedure,
$i$-band sources after deblending ({\tt deblend\_nchild}=0) were used. First we selected all the $i$-band sources within \timeform{4"} from the
position of X-ray sources. 96.5\% of the X-ray sources in the XMM-XXL with EPIC-PN counts $>70$ in 0.2--12 keV have a positional uncertainty smaller than \timeform{4"}. 
Then we examined detection flags for the selected $i$-band sources, and required that all the sources within \timeform{4"} from an X-ray source
are ``cleanly" detected in the $i$ band. 
We used the following criteria for ``clean" detection as used in Toba et al. (2015): 
(1) {\tt flags\_pixed\_edge = not True}, 
(2) {\tt flags\_pixel\_saturated\_center = not True}, 
(3) {\tt flags\_pixel\_cr\_center = not True}, 
(4) {\tt flags\_pixel\_bad = not True}, 
(5) {\tt flags\_cmodel\_flux\_flags = not True}, 
(6) {\tt centroid\_sdss\_flags = not True}, 
(7) {\tt detect\_is\_tract\_inner = True}, and
(8) {\tt detect\_is\_patch\_inner = True}.
Details about the detection flags are given in Aihara et al. (2017b).

At the limiting magnitude of the HSC survey, there are non negligible number of nearby $i$-band sources unrelated to an X-ray source, 
and a simple selection of a source nearest to an X-ray source could result in wrong identifications.
For the X-ray sample matched with 3.6 $\mu$m sources constructed in section 3.2, 32\% of X-ray sources have two or more $i$-band sources within
\timeform{4"} from an X-ray source position.
We use infrared data to better identify the most probable $i$-band counterpart to an X-ray source as described in the next subsection.

\begin{figure}
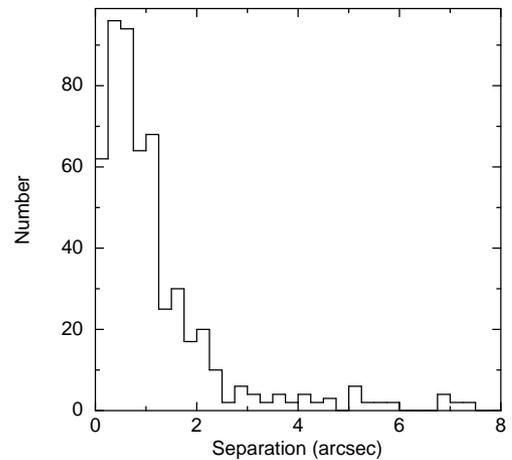

 \begin{center}
\FigureFile(70mm,57mm){fig1.ps} 
 \end{center}
\caption{Distribution of separation between X-ray and its nearest 3.6 $\mu$m sources.}\label{fig:hist_x_ir}
\end{figure}

\subsection{Matching X-ray and infrared sources}

In order to find most probable $i$-band counterparts to X-ray sources,  we use infrared data  because of the following reasons. 
In the optical/infrared identifications of X-ray selected sources using quantitative criteria (likelihood ratios and reliabilities), 
near infrared ($K$ or 3.6 $\mu$m) sources are more likely to be associated with X-ray sources compared to selections using likelihood ratios simply calculated in the optical (Brusa et al. 2007, Akiyama et al. 2015). 
Our main targets, X-ray bright optically faint sources, could be partly missed 
if likelihood ratios in the optical are used since the likelihood selection tends to choose a bright optical source (e.g., Brusa et al.2007). 
Secondary, previous X-ray surveys show that objects with a large X-ray/optical ratio tend to have red infrared/optical colors 
(Mignoli et al. 2004, Brusa et al. 2010).
Thirdly, positional uncertainties  of {\it Spizter} 3.6 $\mu$m sources are much smaller than those for X-rays, and identification processes
between $i$ band and 3.6 $\mu$m are easier than directly matching X-ray and $i$-band sources.

We utilize a catalog of 3.6 $\mu$m sources to first identify infrared counterparts to X-ray sources, and then match infrared 
and $i$-band sources.
The 3.6 $\mu$m band is used because of its best point spread function and sensitivity among the {\it Spitzer}
IRAC (3.6, 4.5, 5.8, 8.0 $\mu$m) and MIPS (24 $\mu$m) bands in the SWIRE survey.
Figure\ref{fig:hist_x_ir} shows a histogram of the distance between X-ray sources and its nearest 3.6 $\mu$m source.
The distances are smaller than \timeform{4"} for 95\% of the X-ray sources.
This distribution is almost identical to that of X-ray positional errors, indicating
the positional uncertainties are dominated by X-ray positional error. We regard infrared sources nearest to 
X-ray sources are infrared counterparts to X-ray sources if the separation is smaller than \timeform{4"}.
There are 432 pairs of X-ray and 3.6 $\mu$m sources after this matching process.
The expected number of an unrelated 3.6 $\mu$m source located within \timeform{4"} from an X-ray source position is 0.12.
The separations between the 3.6 $\mu$m and X-ray positions of 49 out of the final sample consisting of 53 objects (section 3.3) are
less than \timeform{2"}. The expected number of chance coincidence of an infrared source within \timeform{2"} from an X-ray source position 
is 0.03.

\begin{figure}
 \begin{center}
\FigureFile(80mm,50mm){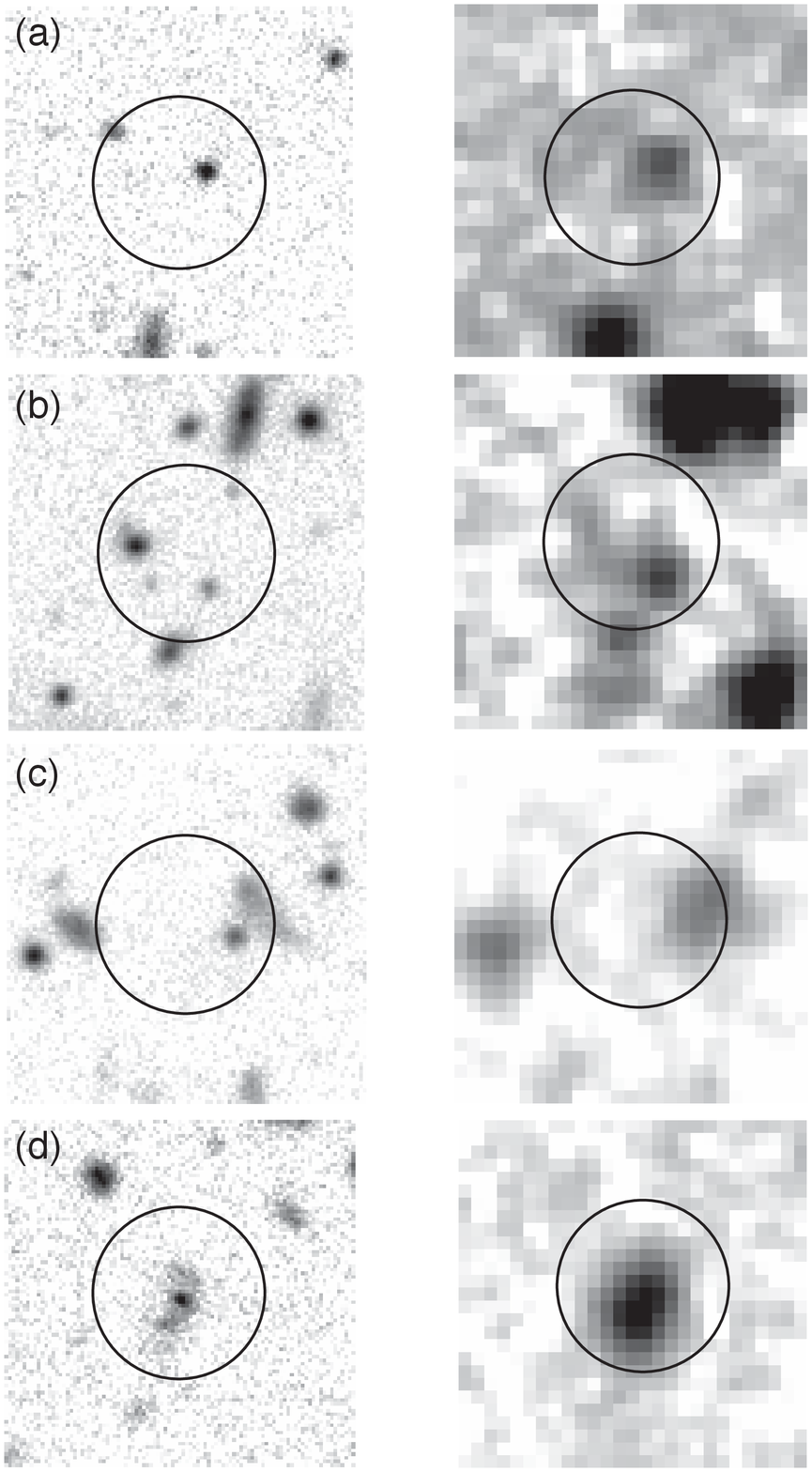} 
 \end{center}
\caption{Examples of $i$ band (left) and 3.6 $\mu$m (right) images.
Image size is 8\timeform{"}$\times$8\timeform{"}.
North is up and east is to the left.
Position of X-ray source is at the center of each panel.
Circle of \timeform{4"} radius from X-ray position is shown.
(a) J021614.5$-$050351.  
(b) J022421.1$-$040355.
(c) J021825.6$-$045945.
(d) J022504.5$-$043706.
}\label{fig:images}
\end{figure}

\subsection{Matching 3.6 $\mu$m and $i$-band sources}

The $i$-band sources nearest to the 3.6 $\mu$m sources selected in section 3.2 are regarded as the
counterparts of the pairs of X-ray and 3.6 $\mu$m sources. 
If $i$-band magnitude of a selected source is brighter than 23.5, they are excluded from our sample. 
77 objects are selected after this screening. 
We examined their X-ray images and found that 24 among 77 objects are not clearly visible in X-ray images
or X-ray counts are smaller than 70 after data screening.  After excluding these sources,
53 X-ray sources with 3.6 $\mu$m and $i$-band counterparts are selected as the final sample.

Examples of $i$-band and 3.6 $\mu$m images are shown in figure\ref{fig:images}.
There is only one $i$ band and 3.6 $\mu$m sources near the X-ray position of 3XMM~J021614.5$-$050351.
In the $i$-band image around 3XMM~J022421.1$-$040355, 
there are three $i$-band sources at a similar distance from the X-ray source, but there is one bright source in 3.6 $\mu$m.
The $i$ band source nearest to this 3.6 $\mu$m source is regarded as the true counterpart.

In three cases (3XMM~J021744.1$-$034531, 3XMM~J021825.6$-$045945, and 3XMM~J022504.5$-$043706), 
there are two $i$-band sources at similar distances from an infrared source, i.e., 
the difference of the separations is less than \timeform{0".5}. 
We examined their $i$-band images and found that 
an infrared peak is located in the middle of two $i$-band sources.
We tentatively assign the $i$-band source nearest to 3.6 $\mu$m source as a counterpart.
Note that such an infrared source might be a combination of emission from the two $i$-band sources, 
or from an interface between two $i$ band sources (e.g., interacting galaxies). 
$i$-band and 3.6 $\mu$m images of the latter two objects are shown in
figure\ref{fig:images} as examples. 
The full set of $i$ band and 3.6 $\mu$m images is presented in the Appendix of the electronic version of this paper.

\subsection{Matching 3.6 $\mu$m and $J, H, K$ band sources}

In order to better constrain the spectral energy distributions (SEDs), near infrared catalogs obtained in
the UKIDSS are matched with our sample. We selected the $K$  band source nearest to the position of 3.6 $\mu$m sources as a counterpart, 
and then $J$ (and $H$ when availble) band photometry for the same sources are compiled.
Among the 53 sources in our sample, 24 and 15 objects are detected in at least one near infrared band in the DXS and 
UDS fields, respectively,  in the UKIDSS.

\begin{figure}
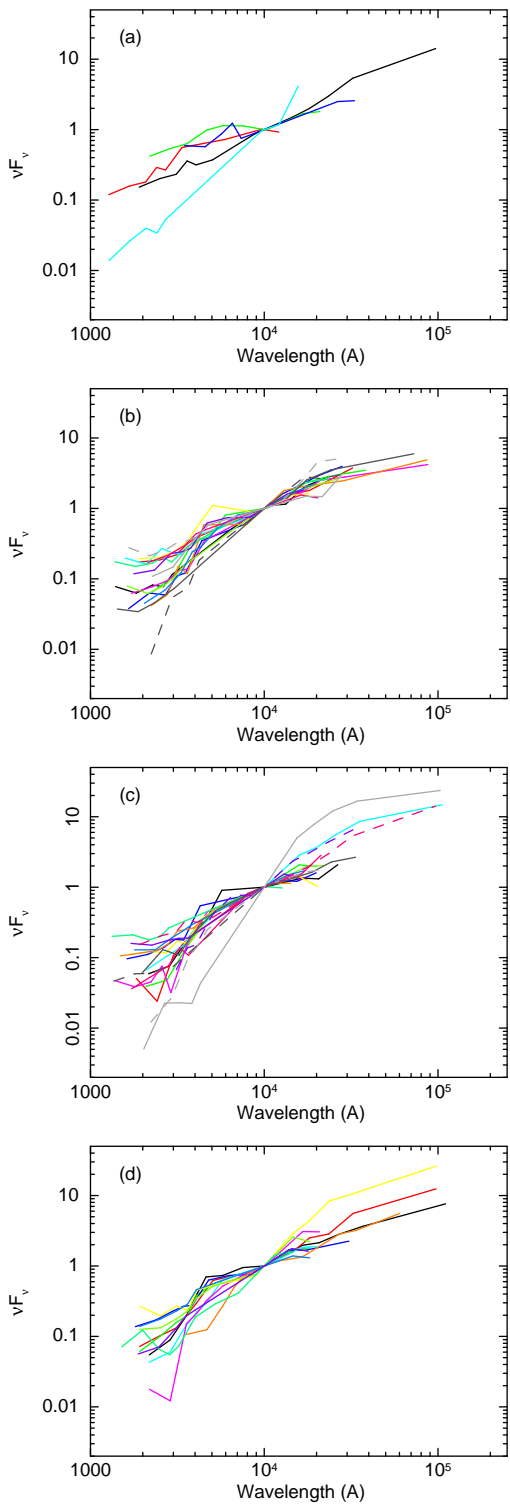

 \begin{center}
 \FigureFile(70mm,57mm){fig3a.ps} 
 \FigureFile(70mm,57mm){fig3b.ps} 
 \FigureFile(70mm,57mm){fig3c.ps} 
 \FigureFile(70mm,57mm){fig3d.ps} 
 \end{center}
\caption{Spectral energy distribution in the source rest frame sorted by template classes used in fits. 
Photometric redshifts determined by SED fits are assumed. Fluxes are normalized at 1 $\mu$m.
(a) AGN1 (b) Mrk 231, (c) AGN2  excluding Mrk 231, and (d) SF. Each of the objects is represented by a different combination of color and line style.
}\label{fig:sed}
\end{figure}

\section{Spectral energy distribution and photometric redshift}

Using the results of multiwavelength matching described in the previous section, we constructed SEDs in the optical and infrared bands for our sample.
We performed SED fits using templates of various types of galaxies 
 and obtained photometric redshifts.
The HYPERZ code (Bolzonella et al. 2000) was used to perform chi-square minimization fits. 
We used the templates compiled by Polletta et al. (2007). 
The templates are categorized into the three broad classes type 1 AGN (AGN1), type 2 AGN (AGN2), 
and starforming galaxy (SF).
The AGN1 class contains the three templates BQSO1, QSO1, TQSO1, 
which are templates of type 1 quasars with different values of infrared/optical ratio.
Their infrared/optical ratios become large in this order.
There are nine templates in the AGN2 class; 
IRAS 19254$-$7245 South (I19254), 
IRAS 20551$-$4250 (I20551), 
IRAS 22491$-$1808 (I22491), 
heavily absorbed broad absorption line quasar (Mrk 231), 
NGC 6240 (N6240), type 2 quasar (QSO2), Seyfert 1.8 (Sey18), 
Seyfert 2 (Sey2), and 
Torus. The three IRAS galaxies (I19254, I20551, I22491), Mrk 231, and NGC 6240 
among them contain a powerful starburst component in addition to AGN.
Templates of seven normal spiral/lenticular galaxies (S0, Sa, Sb, Sc, Sd, Sdm, Spi4) and three starburst templates
Arp 220, M82, and NGC 6090 (N6090) are in the SF class.
Templates of elliptical galaxies were also examined.
In order to represent the difference of extinction among objects in our sample and these templates,  
additional extinction up to $A_V$ = 2.0 was allowed in the fits unless otherwise noted, and 
the reddening law of Calzetti et al. (2000) was assumed.

All the available bands were first used in the fits. Our template fits provided good description of the overall SED for 34 among 53 objects. 
The results of these fits are summarized in table \ref{table:sed}.
We inspected the fitting results and found that in the rest of the cases (19 objects) the model giving $\chi^2$ minimum does not adequately describe the data particularly
around a spectral break seen in 8000--11000 {\AA} in the observed frame.
In order to fit such structures, 
we first fitted their SEDs using $K$ and shorter bands to better constrain its photometric redshift using the spectral break for objects with available $K$ band photometry.
The optical to near infrared SEDs up to $K$ band in 9 objects are described by these fits. 
The model SEDs underpredict infrared excess in the {\it Spitzer} bands except for one object (J021410.3$-$040224), 
indicating the presence of strong emission from dust in addition to the component described by the template.
The SED of J022410.3.0$-$040224 is described by the TQSO1 template up to the $K$ band if extinction of $A_V$=2.3 is allowed. This model slightly over predict infrared fluxes 
at 4.5 and 5.8 $\mu$m by 0.2 dex.
The fit results using SED up to $K$ band are shown in table \ref{table:sed}, where objects showing infrared excess are denoted as ``template name + IR''.

In the rest of the objects, the optical-NIR features were not satisfactory described by above trials. 
Six objects are fitted by restricting the redshift range to $z = 1-2$, in which local minima of $\chi^2$ are obtained, using full band SED.
A model giving a local minimum in the redshift range $z=2-3$ describes the SED of one object  (J022331.0$-$044234).
One object (J021634.3$-$050724) is fitted by restricting $z$ to $1-2$ and using SED up to $K$ band.
The overall spectral shape and the break features in two objects are explained by using the template of Mrk 231 (J021744.1-034531 and J022330.8-044632).
These results are also summarized in table \ref{table:sed}.

The templates used in the SED fits are classified into three classes AGN1, AGN2, or SF.
These classifications are also shown in table \ref{table:sed}.
The results show that only a very small fraction (5 out of 53) are classified as AGN1 and most of objects are explained by AGN2 or SF templates.
None are represented by templates of early type galaxies.
figure \ref{fig:sed} shows the normalized SEDs in the source rest frame adopting the photometric redshifts determined from the fits. 
The availble data points are connected by a solid line for each source.
In figure \ref{fig:sed}, SEDs are divided into four groups by templates; AGN1, 
Mrk 231, AGN2 excluding Mrk 231, and SF. The group of Mrk 231 is separately shown for clarity because a large number of 
objects (13) are fitted by this template.
The SEDs of AGN1s in our sample do not show prominent spectral features except for a small bump in one band that might be due to contribution of emission lines.
Therefore, the photometric redshifts for these objects could have large uncertainties. 
The optical part of their SEDs is relatively red, implying reddening by dust.
There is a break feature at around 4000 {\AA} in  SEDs of AGN2 and SF except for a few objects showing relatively featureless continuum.
This feature is weak but present in most of the objects in the Mrk 231 group. 
The photometric redshifts are primarily determined by this feature in combination with the overall SED shape.

\begin{figure}
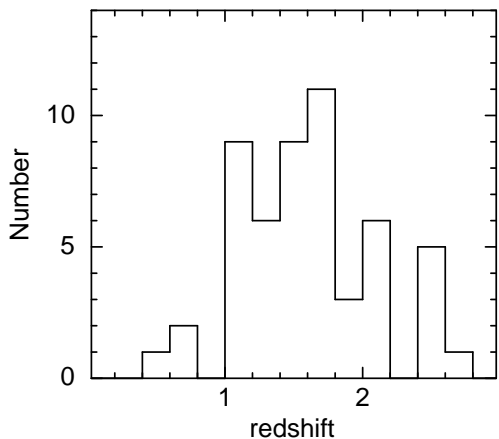

 \begin{center}
\FigureFile(70mm,57mm){fig4.ps} 
 \end{center}
\caption{Distribution of redshifts.
Spectroscopic redsfhifts for seven objects and photometric redsfhits for the rest are used.
}\label{fig:hist-z}
\end{figure}

Results of spectral analysis of seven objects in our sample are presented in Akiyama et al. (2015), and  their measurements of 
spectroscopic redshifts as well as spectral classifications are also shown in table \ref{table:sed}. 
These spectroscopic redshifts are assumed if available, and the photometric redshifts we derived 
from the SED fits are used otherwise in the following analysis. 
The distribution of the redshifts are shown in Figure \ref{fig:hist-z}. three (5\%), 38 (72\%), and 12 (23\%) of the sample 
are located at $z<1$, 1--2, and 2--3, respectively.

\section{X-ray Spectra}

We made X-ray spectra of the 53 objects in our sample. The observation log is  shown in table \ref{table:obslog}.
X-ray spectra were extracted from a circular region with a radius of 200--300 pixels or 10--\timeform{15"}. 
The extraction radius was chosen based on the brightness of an X-ray source and to avoid nearby X-ray sources, if any.  
Background spectra were extracted from an off-source region in the same CCD chip. 
The high background regions near edges of the CCD chips were not used.
In some cases, the source position is out of the field of view of one or two sensors, and only available data were used in the analysis. 
Such cases are shown in the footnotes of table \ref{table:pl}.
Event PATTERN of $\leq 12$ and $\leq4$ were used for EPIC-MOS and EPIC-PN spectra, respectively.
The total net counts in the 0.5--10 keV band from all the available sensors are shown in table 
\ref{table:pl}.
The response matrix files (RMFs) and ancillary response files (ARFs) were created by using {\tt rmfgen} and {\tt arfgen} in the SAS package.
The X-ray spectra were analyzed using  XSPEC version12.9.0n.
Since the photon statistics are relatively poor for all the objects, the spectra were binned so that each bin contains at least one count, 
and fitted by maximum likelihood method using $C$ statistic (Cash 1979).
The errors represent the 90\% confidence interval for one parameter of interest.
The Galactic absorption were applied to all the models examined below.
The Galactic column densities are fixed at the values obtained by 
the {\tt nh} tool in the HEASOFT6.19 according to the H\emissiontype{I} map by Kalberla et al. (2005).
The used $ N_{\rm H}$ values are tabulated in table \ref{table:pl}.

\subsection{Apparent spectral slope}

\begin{figure}
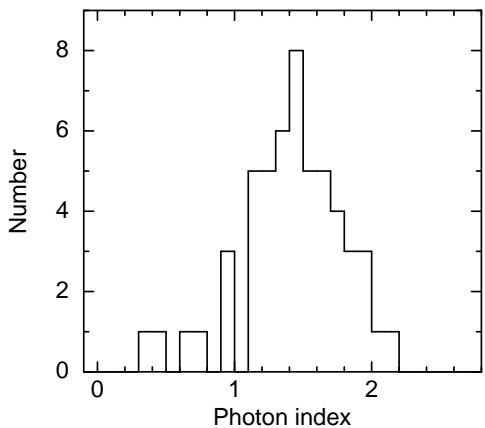

 \begin{center}
 \FigureFile(68mm,55mm){fig5.ps} 
 \end{center}
\caption{Distribution of apparent photon indices.}\label{fig:index}
\end{figure}

\begin{figure}
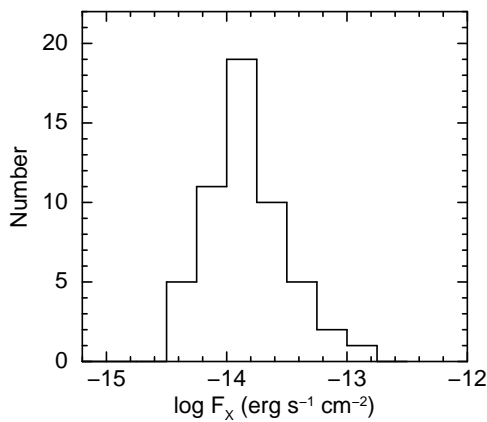

\begin{center}
\FigureFile(70mm,57mm){fig6.ps} 
 \end{center}
\caption{Distribution of observed X-ray fluxes in 2--10 keV.}\label{fig:Fx}
\end{figure}

We first measured apparent spectral slopes without assuming the source redshifts.
A power law model absorbed only by Galactic $N_{\rm H}$ was assumed. The free parameters are a photon index ($\Gamma$) 
and a normalization of the power law.
The results are shown in table \ref{table:pl}. The apparent photon indices (figure \ref{fig:index}) are distributed around $\Gamma \approx 1.5$, which is slightly flatter than those observed in unobscured Seyferts or quasars ($\Gamma \approx1.7-1.9$; e.g., Nandra \& Pounds 1994, Piconcelli et al. 2005) 
or their  intrinsic photon index ($\Gamma \approx1.9$; e.g., Nandra et al. 2007). 
Only eight objects show very flat best-fit slopes ($\Gamma<1.0$). 
The observed flux in the 2--10 keV band is summarized in table \ref{table:pl} and figure \ref{fig:Fx}
The fluxes are in the range of $6.0\times10^{-15} - 1.7\times10^{-13}$ erg s$^{-1}$ cm$^{-2}$.

\subsection{Absorption column density}

We next fitted the X-ray spectra by an absorbed power-law model to constrain the amount of absorption. 
An intrinsic absorber is assumed to be located at the 
photometric redshift determined in section 4 or spectroscopic redshift when available.
Since the photon statistics are limited, the photon index is fixed at $\Gamma=1.9$. 
The free parameters are an intrinsic absorption column density and a normalization of the power-law component. 
This model provides a good description of the X-ray spectra.
One object, J021529.4$-$034322 (figure8 (a)), shows weak wavy residuals peaking at 0.5 keV and 2 keV. If an unabsorbed power-law component is added,
a column density of $1.6^{+6.2}_{-1.2}\times10^{23}$ cm$^{-2}$ for an absorbed component
and an only  slightly better value of $C$  statistic ($\Delta C = -4.7$) 
are obtained for one additional free parameter (normalization of unabsorbed power law). 
In the following analysis, we use the results of single absorbed power law fits for all the objects.
The observed spectra, best-fit absorbed power law model, and data/model ratios for objects with net EPIC-PN counts in 0.5-10 keV
greater than 90 (12) objects in total are shown in figure \ref{fig:xspec}.
The spectra taken with EPIC-MOS1 and EPIC-MOS2 are combined for presentation purpose.

The intrinsic absorption column densities thus obtained are tabulated in table \ref{table:abs} 
and their distribution is shown in figure \ref{fig:nh} as a histogram. 
The best-fit absorption column densities are distributed in the range of $\log N_{\rm H} =20.5 - 23.5$ (cm$^{-2}$), 
except for three objects, for which  the best-fit value becomes zero.
The column densities of most objects are modest, as implied from the apparent slope in the single power-law fits in section 5.1,  
with a distribution peaking at $\log N_{\rm H} \approx 21.5-22$ (cm$^{-2}$).
Note that the derived absorption column densities depend on the assumed redshift. The dependence is approximated by
$N_{\rm H} \approx (1+z)^{2.5} N_{{\rm H}, z=0}$ for redshifts greater than $\sim$ 0.5, where $N_{{\rm H}, z=0}$ is the absorption column density
obtained by assuming $z=0$ in spectral fits.

By using the absorbed power-law model,
intrinsic X-ray luminosities in the 2--10 keV band (source rest frame) corrected for both Galactic and intrinsic 
absorption were derived, where photometric or spectroscopic redshifts were assumed. 
The intrinsic luminosities are  shown in table \ref{table:abs} and figure \ref{fig:Lx}. 
The peak of the distribution of the intrinsic luminosities is around $\log L_{\rm X}$=44.0-44.4 erg s$^{-1}$. 
The boundary between Seyfert and quasar luminosities in X-rays of $10^{44}$ erg s$^{-1}$ in 2--10 keV is often used, 
and luminosities of most of objects in our sample are those of quasars or luminous Seyferts. 
Among 35 objects having $ L_{\rm X}>10^{44}$ erg s$^{-1}$, $N_{\rm H}$ of 20 objects are in excess of $10^{22}$ cm$^{-2}$ 
and are classified as type 2 quasars based on X-ray absorption. Thus our selection is efficient to find a population of type 2 quasars.
Among the 20 type 2 quasar candidates, the redshift of one object (J021736.4$-$050106) was spectroscopically measured
(table 2; Akiyama et al. 2015).

\begin{figure}
\begin{center}
\FigureFile(62mm,45mm){fig7a.ps} 
\FigureFile(62mm,45mm){fig7b.ps} 
\FigureFile(62mm,45mm){fig7c.ps} 
\FigureFile(62mm,45mm){fig7d.ps} 
  \end{center}
\caption{
X-ray spectra fitted by an absorbed power law model. The observed data are binned for presentation purpose.
Crosses with filled circle: EPIC-PN, Crosses without circle: combined EPIC-MOS1 and EPIC-MOS2.
(Upper panel) Data and model. Upper solid histogram: best-fit model for EPIC-PN; lower histogram: best-fit model for EPIC-MOS.
(Lower panel) Data/Model.
(a) J021529.4$-$034233.
(b) J021625.7$-$050518.
(c) J021634.3$-$050724.
(d) J021642.3$-$043552.
}
\label{fig:xspec}
\end{figure}

 \addtocounter{figure}{-1}
 \begin{figure}
 \begin{center}
\FigureFile(62mm,45mm){fig7e.ps} 
\FigureFile(62mm,45mm){fig7f.ps} 
\FigureFile(62mm,45mm){fig7g.ps} 
\FigureFile(62mm,45mm){fig7h.ps} 
 \end{center}
\caption{
Continued.
(e) J021842.8$-$051934.
(f) J022145.5$-$034346.
(g) J022214.4$-$034619.
(h) J022330.8$-$044632.
}
\label{fig:xspec}
\end{figure}

\addtocounter{figure}{-1}
 \begin{figure}
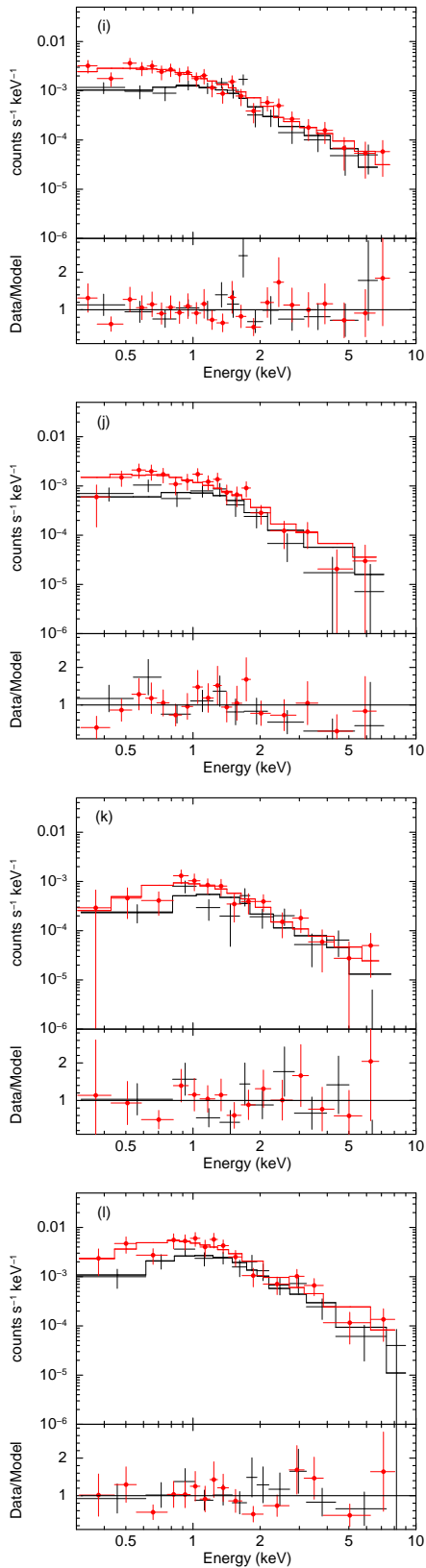

 \begin{center}
\FigureFile(62mm,45mm){fig7i.ps} 
\FigureFile(62mm,45mm){fig7j.ps} 
\FigureFile(62mm,45mm){fig7k.ps} 
\FigureFile(62mm,45mm){fig7l.ps} 
 \end{center}
\caption{
Continued.
(i) J022405.2$-$041612.
(j) J022420.7$-$041224.
(k) J022421.1$-$040355.
(l) J022510.6$-$043549.
}
\label{fig:xspec}
\end{figure}

\begin{figure}
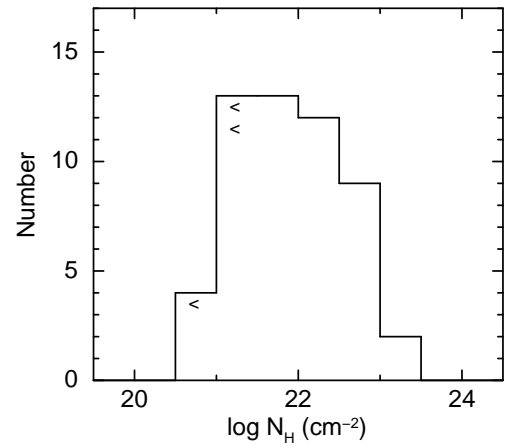

 \begin{center}
 \FigureFile(70mm,57mm){fig8.ps} 
 \end{center}
\caption{Distribution of best-fit values of absorption column density. Photon index is fixed at 1.9 in the fits. 
Photometric or spectroscopic (when available) redshifts are assumed in the spectral fits.}
"$<$" symbol denotes upper limit
for objects with best-fit column density of zero.
\label{fig:nh}
\end{figure}

\begin{figure}
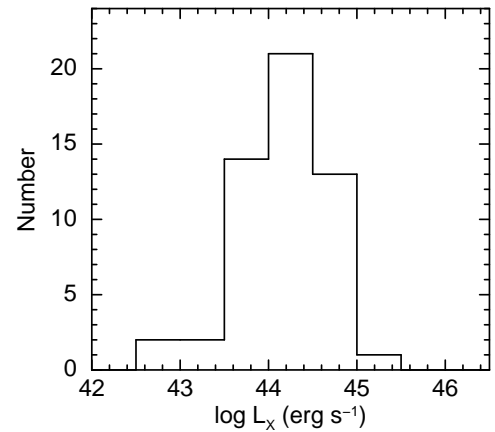

 \begin{center}
 \FigureFile(68mm,55mm){fig9.ps} 
 \end{center}
\caption{Distribution of X-ray luminosities in 2--10 keV corrected for absorption.
Photometric or spectroscopic (when available) redshifts are assumed.
}\label{fig:Lx}
\end{figure}

\begin{figure}
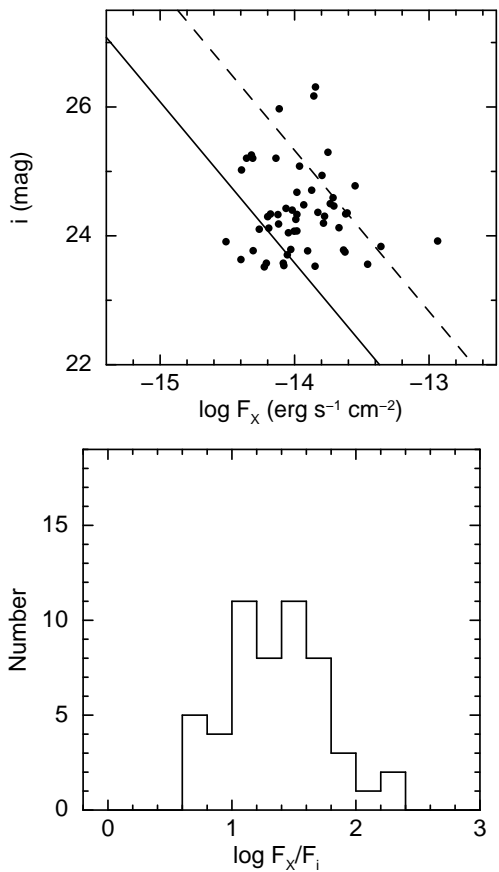

 \begin{center}
  \FigureFile(70mm,57mm){fig10a.ps} 
  \FigureFile(70mm,57mm){fig10b.ps} 
 \end{center}
\caption{
(Left panel) $i$-band magnitude corrected for Galactic extinction plotted against $\log F_{\rm X}$ (erg s$^{-1}$ cm$^{-2}$) corrected for absorption in the 2--10 keV band.
Solid and dashed lines correspond to flux ratios of $F_{\rm X}/F_i$ = 10 and 50, respectively.
(Right panel) Distribution of X-ray flux to $i$-band flux ratio $F_{\rm X}/F_i$.
X-ray fluxes are in the 2--10 keV band corrected for the Galactic absorption. 
$i$-band fluxes are corrected for the Galactic extinction.
}\label{fig:FxFo}
\end{figure}

\subsection{Flux ratios}

The X-ray fluxes in the 2--10 keV band corrected for the Galactic absorption derived from the absorbed power-law fits ($F_{\rm X}$)
are used to calculate X-ray to $i$-band flux ratios. 
$i$-band flux $F_i$ is defined as $\Delta \lambda f_\lambda$, where $\Delta \lambda$ is 
a full width at half maximum of the $i$-band filter transmission (1480 {\AA})
and $f_\lambda$ is a flux density at $i$ band in units of erg s$^{-1}$ cm$^{-2}$ \AA$^{-1}$.

The $i$-band magnitude corrected for the Galactic extinction, X-ray flux in the 2--10 keV band corrected for the Galactic absorption, 
and the distribution of the flux ratios are shown in figure\ref{fig:FxFo}. 
The values of flux ratios are summarized in table \ref{table:ratio}.
The $F_{\rm X}/F_i$ ratios for 44
objects are greater than 10, which are classified as extreme X-ray-to-optical flux sources (EXOs). 9 among them have $F_{\rm X}/F_i>50$ 
and are classified as EXO50 that are a class showing most extreme X-ray to optical flux ratios as defined in Della Ceca et al. (2015) using the $R$ band.
The deficit of objects with small 
$F_{\rm X}/F_i$ ratios are due to our selection criteria;  we selected X-ray bright (0.2--12 keV EPIC-PN counts greater than 70) 
and optically faint ($i$-band magnitude fainter than 23.5) objects.

\begin{figure}
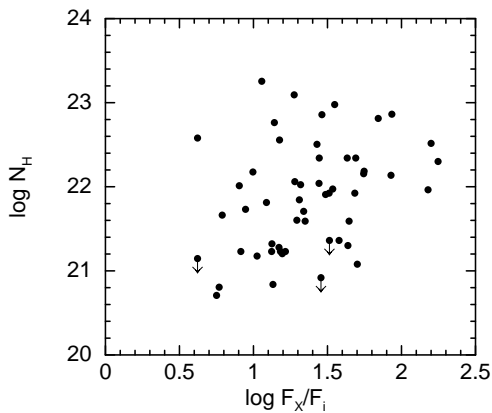

 \begin{center}
 \FigureFile(70mm,57mm){fig11.ps} 
 \end{center}
\caption{
Absorption column density and X-ray flux to $i$-band flux ratio $F_{\rm X}/F_i$.
}
\label{fig:FxFo-nh}
\end{figure}

\section{Discussion}

\subsection{X-ray absorption}

We made a sample of 53 X-ray bright optically faint AGNs and analyzed their X-ray spectra. 
The number of objects for which individual X-ray spectral fit can be performed is 
much larger than previous works (Perola et al. 2004, Rovilos et al. 2010, Dela Ceca et al. 2015).
X-ray spectral fits with an absorbed power-law model show that the X-ray absorption is modest; 43 among 53 objects are absorbed by a column density 
between $10^{21}$ and $10^{23}$ cm$^{-2}$. Only three are absorbed by the best-fit column of an order of $10^{23}$ cm$^{-2}$. Thus all the objects are classified as Compton-thin AGNs. We examine whether flat X-ray spectra observed in several objects are explained by a Compton-thick AGN. All of the X-ray spectra of seven sources with an apparent photon index of $\Gamma<1.0$ show a clear convex shape peaking at energies $1.2-2$ keV indicating mild absorption. 
This shape is not compatible with a reflection dominated X-ray spectrum that is flat, 
nor a combination of reflected emission and emission Thomson scattered by ionized medium resulting a concave shape. 
We therefore conclude that our spectra are best explained by a power law model absorbed by at most several $\times 10^{23}$ cm$^{-2}$
as deduced from absorbed power law fits in section 5.2.

The modest absorption measured for our sample and absence of Compton-thick AGN are in agreement with results obtained with {\it XMM-Newton} using a much smaller survey area.
The observed 2--10 keV fluxes for our sample are in the range of
 $F_{\rm X}=4\times10^{-15} - 2\times10^{-13}$ erg s$^{-1}$ cm$^{-2}$.
This range covers X-ray fainter sources compared to objects found in the HELLAS2XMM survey and XMDS by a factor of two (Perola et al. 2004, Tajer et al. 2007).
Therefore, we found no evidence for emergence of new properties in the covered flux range.
On the other hand, previous studies of X-ray bright optically faint AGNs using {\it Chandra} show a larger fraction of objects with flatter slope than those for our sample.
This discrepancy might be partly due to the flux range.
In {\it Chandra} studies using deep fields,
fainter sources (mostly $<$ several $\times10^{-15}$ erg s$^{-1}$ cm$^{-2}$; e.g., Civano et al. 2005) are sampled,
and heavily absorbed AGNs with a similar luminosity could be detected. 
Another possible selection bias is our use of full band X-ray counts in the sample selection. 
The effective area  of {\it XMM-Newton} is larger at soft X-rays and more sensitive to
less absorbed objects.

Next we test whether the relation between the optical faintness of our sample is related to the X-ray absorption ($N_{\rm H}$). 
As shown in figure \ref{fig:FxFo-nh}, a correlation
between X-ray to $i$-band flux ratio ($F_{\rm X}/F_i$)  and $N_{\rm H}$ is not clear. 
Hereafter, we use X-ray fluxes corrected for the Galactic absorption derived from the absorbed power law fits and
$i$-band fluxes corrected for the Galactic extinction. 
All of the $N_{\rm H}$ value of most extreme objects with $\log (F_{\rm X}/F_i)
 > 1.9$ are greater than $10^{22}$ cm$^{-2}$, 
although it is not clear whether this trend is real or due to the limited sample size.

X-ray absorption column densities obtained for our sample are mostly in the range $\log N_{\rm H} \approx 20.5-23.5$ cm$^{-2}$, 
which corresponds to an extinction of $E_{B-V}$=0.054--54 or $A_V$ = 0.17--170 
if standard conversion factors of $A_V/E_{B-V}$ = 3.1 and $N_{\rm H}/E_{B-V} = 5.8\times10^{21}$ cm$^{-2}$ mag$^{-1}$
(Bohlin et al. 1978) are assumed.
 We examine whether this amount of absorption is sufficient to obscure the central AGN in the optical and 
to explain the optical faintness of our sample. If the intrinsic SED between optical and X-ray is represented by a template of unabsorbed radio quiet quasars 
assembled by Elvis et al. (1994),
the slope between the $V$ band and 1 keV is $\log \nu F_{\nu, V} - \log \nu F_{\nu , \rm 1 keV} \approx 0.6$. 
The flux ratios of $F_{\rm X}/F_i$ = 10 and 50 correspond to $A_V$ = 1.5 and 3.3 if a quasar is the only source of radiation. 
The $N_{\rm H}$ values required for these $A_V$  values are $2.8\times10^{21}$ and $6.2\times10^{21}$ cm$^{-2}$, respectively.
The best-fit $N_{\rm H}$ values of 15 objects do not exceed the former value and the optical faintness cannot be explained if the standard conversion factor is assumed.
The SEDs of most objects show signature of stellar emission component in the optical band in the source rest frame, 
the extinction necessary to hide a large fraction of AGN emission should be higher than these $A_V$ values. 
Therefore, $E_{B-V}/N_{\rm H}$ values are suggested to be larger than the standard value for at least a part of the sample.
This condition implies a large dust/gas ratio in galaxies in our sample and then smaller $N_{\rm H}$ could give stronger reddening.
This dust rich nature of our sample is consistent with the significant infrared emission seen in the SED.
Note, however, that $E_{B-V}/N_{\rm H}$ values in AGNs are suggested to be smaller than the  Galactic standard value (Maiolino et al. 2001), which is in the opposite sense to our case.
Their sample consists of Seyfert galaxies showing at least two broad lines in optical/infrared with $N_{\rm H}$ measured by X-ray, 
which are ordinary AGNs in terms of 
X-ray to optical flux ratios. Thus the dust rich nature is suggested to be related to optical faintness in our sample.

The peak of the number density of AGNs of an X-ray luminosity of $10^{44-45}$ erg s$^{-1}$, which is typical for our sample, is at around $z\sim 1.6$ (Ueda et al. 2014). 
Redshifts for most of objects in our sample (50 objects) are located at a redshift in a range between 1.0 and 2.7, which coincides with the peak of the number density.
The observed modest X-ray absorption column densities indicates that they are not missed in surveys at $<$10 keV  (or $<$30 keV at $z=2$). On the other hand,
the optical faintness means relatively deep surveys are necessary to fully sample 
this class of objects in the optical.

\begin{figure}
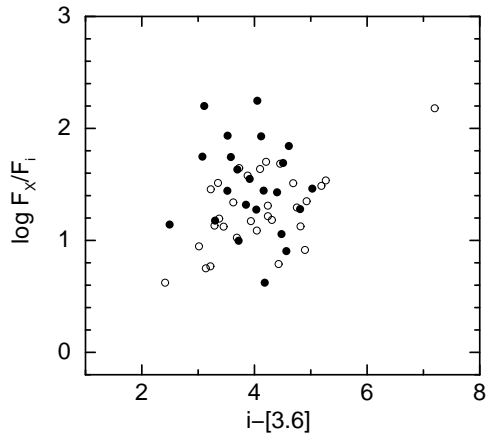

 \begin{center}
 \FigureFile(70mm,57mm){fig12.ps} 
 \end{center}
\caption{$\log (F_{\rm X}/F_i)$ plotted against $i$-[3.6] color in AB magnitude.
Open circles: $\log N_{\rm H} <22$ cm$^{-2}$. Filled Circles: $\log N_{\rm H} >22$ cm$^{-2}$. 
}\label{fig:FxFoi36}
\end{figure}

\begin{figure}
 \begin{center}
 \FigureFile(70mm,57mm){fig13.ps} 
 \end{center}
\caption{$\log (F_{\rm X}/F_i)$ plotted against [3.6]$-$[4.5] color in AB magnitude.
Open circles: $\log N_{\rm H} <22$ cm$^{-2}$. Filled Circles: $\log N_{\rm H} >22$ cm$^{-2}$. 
}\label{fig:FxFo3645}
\end{figure}

\begin{figure}
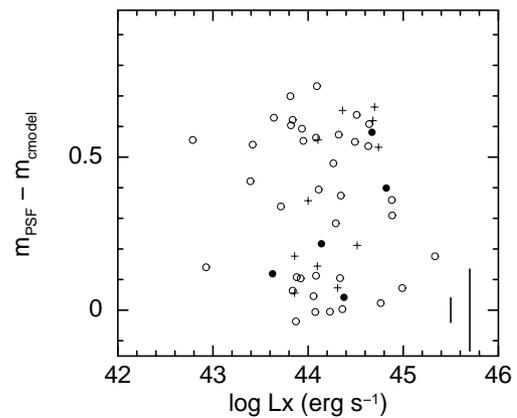

 \begin{center}
\FigureFile(72mm,58mm){fig14.ps} 
 \end{center}
\caption{Difference between  PSF and cmodel magnitudes plotted against absorption corrected 2--10 keV luminosity.
Short and long bars in lower right corner are typical error of $m_{\rm PSF}-m_{\rm cmodel}$ for $i$=24 and 25, respectively.
Filled circles: AGN 1; Open circles: AGN 2; Crosses: SF.
}\label{fig:psf}
\end{figure}

\subsection{Optical/Infrared Color, SED, and Optical Faintness}

Figure \ref{fig:FxFoi36} plots $\log (F_{\rm X}/F_i)$ against $i-$3.6 $\mu$m color ($i-[3.6]$ in AB magnitude).
The $i-[3.6]$ colors are shown in table \ref{table:ratio}.
The locus occupied by the objects in our sample is 
in accordance with the known trend that optically faint X-ray bright objects have redder color between 
the near infrared and optical bands (e.g., $R-K$, $R-[3.6]$; Brusa et al. 2010).
The $i-[3.6]$ colors are in the range of 2.7--7.5 indicating very red color and faintness in the $i$ band compared to 3.6 $\mu$m. 
Open and filled symbols in figure\ref{fig:FxFoi36} are objects with $\log N_{\rm H}<22$, and $>22$ cm$^{-2}$, respectively. 
There is no clear evidence that these groups of small and large absorption show different $i-[3.6]$ colors.

All the objects in our sample are detected at both 3.6 $\mu$m and 4.5 $\mu$m bands. Figure \ref{fig:FxFo3645} shows $\log (F_{\rm X}/F_i)$  plotted against [3.6]$-$[4.5] color (table \ref{table:ratio}). 
The [3.6]$-$[4.5] colors are also red (0.05--0.82) compared to the distribution for X-ray selected AGNs containing both broad-line and non-broad-line AGNs. 
(e.g., [3.6]$-$[4.5] = $-0.8$~--~$+0.9$, Feruglio et al.  2008). The observed red color in the near infrared band in the source rest frame
indicates the presence of hot dust most probably heated by AGN.

The results of SED fits indicate the presence of significant amount of dust in most of the objects in our sample.
Only five objects are fitted by type 1 AGN template, and the rest of the objects are represented by type 2 AGN or SF galaxy. 
The $A_V$ values of the five objects showing Type 1 SED are in the range of $1.1-2.3$ leading to a relatively red SED in the optical
and a large $F_{\rm X}/F_i$ ratio.
Furthermore, most objects show significant infrared emission;  a large fraction of the sample (44 of 53) are explained by SED 
of ultraluminous infrared galaxies (ULIRGs), starburst galaxy, AGN with emission from dusty torus, or type 2 AGNs, 
in which rest-UV and optical emission are reddened by dust. 
In the cases that normal galaxy templates are applied to the optical and near infrared SED up to $K$-band, 
significant infrared excess over the model is seen at 3.6 $\mu$m, 4.5 $\mu$m,  and longer wavelengths (if data is available) in most objects.
These results indicate that objects in our sample show strong infrared emission from AGN and/or star formation activity and the presence of remarkable amount of dust,
and the presence of dust heated by AGN is in agreement with the red color in the near infrared band.
The extinction caused by the dust is thus likely to contribute to the faintness in the optical band.

Another cause of optical faintness is the SED shape of stellar component of host galaxies. 
AGN emission is likely to be attenuated by dust in the rest-UV and optical bands, and host emission can be observed. 
Most of the observed SED show a break feature at 8000$-$11000 {\AA} in the observed frame. 
Such features are well explained by galaxy component seen in the templates except for cases fitted by type 1 AGN template or a few objects showing no clear features.
The determined redshift are greater than 1.0 except for three cases. At $z>1$, the 4000 {\AA} break shifts to
the $i$ or redder bands. Therefore the combination of the break feature and the redshift range is another likely cause of the optical faintness.

The presence of the break feature implies that the central AGN is obscured and that stellar component of host galaxies 
is significantly present in the optical emission.
We examine the contribution of host by comparing PSF and cmodel magnitudes. Figure \ref{fig:psf} shows the difference of 
PSF magnitude and cmodel magnitude in the $i$ band
plotted against absorption corrected X-ray luminosities in the 2--10 keV band in the source rest frame. SED types of 
AGN1, AGN2, and SF are shown as different symbols. 
The figure indicates the presence of extended emission component up to 0.8 mag 
in a large fraction of objects being consistent with a significant contribution of host emission
to the observed SED. There is no clear evidence for a correlation between extendedness and X-ray luminosity, 
implying AGN component is absorbed by dust even in luminous  ($>10^{44}$ erg s$^{-1}$) AGNs.

\subsection{Mid-infrared emission and obscuration of AGN}

Ten objects in our sample are detected at 24 $\mu$m in the SWIRE survey. Four of them have a 24 $\mu$m to $i$-band flux density ratio $f_{24}/f_i$
greater than 1000 (or $i-[24]>7.5$; table \ref{table:ratio})
satisfying the criterion of DOGs, where
we use $i$ band instead of $R$ band used in the original definition (Dey et al. 2008).
$f_{24}/f_i$ ratios for two objects (J021721.9$-$043655 and
J022337.9$-$040512) are greater than 2000 and are classified as extremely dust obscured galaxies (EDOGs). 
The absorption column densities of these EDOGs and two DOGs with $1000 \leq f_{24}/f_i \leq 2000$ (J021642.3$-$043552 and 
J021705.4$-$045655) are modest; $0.92^{+0.55}_{-0.40}$, $3.2^{+1.5}_{-1.1}$, $3.3^{+0.53}_{-0.47}$, and $1.1^{+0.54}_{-0.42}\times10^{22}$ cm$^{-2}$, 
respectively.

Among seven EXO50s studied in Della Ceca et al. (2015), three are classified as EDOGs ($f_{24}/f_R>2000$). 
Their absorption column densities $N_{\rm H}$ are measured to be
$>0.4$, $3.61^{+1.03}_{-0.92}$, and $>0.1\times10^{22}$ cm$^{-2}$. 
Lanzuisi et al. (2009) analyzed {\it Chandra} and {\it XMM-Newton} data of  a sample of 44 EDOGS, 
and at least 21 objects are classified as EXOs ($F_{\rm X}/F_R>10$).
The estimated  $N_{\rm H}$ for the 21 objects are typically in the range of $1\times10^{22}$ -- several$\times10^{23}$  cm$^{-2}$ 
with four upper limits, two lower limits, and one 
candidate Compton thick object ($2.7\times10^{24}$ cm$^{-2}$).
A recent study of 14 X-ray detected DOGs in the {\it Chandra} Deep Field South have shown that only three could be considered as Compton-thick AGN,
and the others are absorbed by at most $8\times10^{23}$ cm$^{-2}$ (Corral et al. 2016). 
These modest absorption and  the paucity of Compton-thick AGNs in X-ray detected DOGs 
are in agreement with the measurement in the present sample.
Note, however, that X-ray emission is extremely extinct even at hard X-rays ($>$20 keV in the source rest frame) if absorption column density is greater than $\sim10^{25}$ cm$^{-2}$ 
(Wilman \& Fabian 1999, Ikeda et al. 2009). Current X-ray surveys are biased against such extreme cases, 
and therefore systematic X-ray follow up observations of optically/infrared-selected DOGs is necessary to fully understand 
the true nature of this population.
A part of X-ray undetected DOGs are indeed suggested to be  heavily obscured quasars (Corral et al. 2016).

If an AGN is energetically dominant, mid-infrared emission predominantly comes from warm/hot dust heated by  AGN, and then a good correlation between
X-ray and mid-infrared luminosity is expected unless X-ray emission is significantly attenuated by absorption
(Gandhi et al. 2009, Mateos et al. 2015, Asmus et al. 2015, Ichikawa et al. 2017). 
According to the correlation between 22 $\mu$m and hard X-ray luminosity in 14--195 keV in Ichikawa et al. (2017), 
an X-ray to mid-infrared luminosity ratio  ($L_{\rm X}/\lambda L_{\lambda, \rm MIR}$) of 0.48 is expected for an X-ray luminosity in the 2--10 keV band of $\log L_{\rm X} = 44$ (erg s$^{-1}$), 
which is typical for our sample with 24 $\mu$m detection. In the conversion between luminosities in 14--195 keV and 2--10 keV, 
an unabsorbed AGN spectrum of  $\Gamma=1.9$ is assumed as in Ichikawa et al. (2017). 
X-ray to mid-infrared luminosity ratios for the ten sources with 24 $\mu$m detection range from 0.56 to 7.2 (table \ref{table:ratio}). 
These values are close to or even higher than the expected ratio. This fact indicates that our absorption correction is reliable and that the X-ray emission is not obscured by Compton-thick matter.

\section{Summary and Conclusions}

We constructed a sample of X-ray bright optically faint AGNs by combining Subaru HSC, {\it XMM-Newton}, {\it Spitzer} and UKIDSS surveys.
From a parent sample of $i>23.5$ mag and X-ray counts with EPIC-PN in 0.2--12 keV $>$70 in the part of the XXL survey area covered by {\it Spitzer} public catalog,
53 X-ray bright optically faint AGNs
are selected by matching X-ray, 3.6 $\mu$m, and $i$-band sources. 
The wide area used in this study (9.1 deg$^2$) enabled us to create a large sample for which X-ray spectral analysis is possible. 
Optical/infrared SEDs were created and fitted by templates of various types of galaxies, then 
photometric redshifts were determined.
The nature of the selected sources were investigated by using SEDs, X-ray spectra, optical/infrared colors. Their properties are summarized as follows.\\

\noindent
1. $F_{\rm X}/F_i$ ratios of 44 objects among 53 sources are greater than 10 and classified as extreme-X-ray-to-optical flux sources (EXOs). 
9 among them have $F_{\rm X}/F_i  >50$ that are EXO50s.\\

\noindent
2. SEDs are represented by AGN2 or SF templates, except for five objects fitted by AGN1. Most objects show a break feature at around 8000--11000 {\AA} 
in the observed frame that is likely to be from stellar component of the host galaxies. Red optical/infrared colors suggest that the central AGN is dust obscured.
The determined redshifts (mostly $z=1.0-2.2$) coincide with the peak of the number density of AGNs of X-ray luminosity $10^{44-45}$ erg s$^{-1}$.\\

\noindent
3. X-ray spectra were well represented by an absorbed power law model with a fixed photon index of 1.9. The best-fit intrinsic absorption column densities at the source redshift
are mostly in the range of $\log N_{\rm H} = 20.5-23.5$ cm$^{-2}$. These modest absorption column densities are in agreement with studies of similar populations (EXOs and DOGs). The measured $N_{\rm H}$ values are not always sufficient to obscure AGN emission in the optical band to explain the SED shape. A large dust/gas ratio would be required in a part of our sample.\\

\noindent
4. 20 objects among 53 (38\%) are classified as type 2 quasars. This fraction indicates that our selection criteria efficiently sample type 2 quasars.\\

\noindent
5. Among 10 objects with 24 $\mu$m detection, four are classified as DOGs defined as $f_{24}/f_i>$1000.  Two objects are EDOGs with $f_{24}/f_i>$2000. 
X-ray to 24 $\mu$m ratios implies that our X-ray sources are not obscured by Compton-thick matter and that our absorption corrections are reliable.\\

\noindent
6. The most likely causes of optical faintness in our sample is a combination of spectral break of the stellar component redshifted to the band redder than the $i$ band (i.e., $z>1$)
(except for several objects with relatively featureless SEDs), significant dust extinction in the rest frame UV and optical bands, and in part a large ratio of dust/gas.


\begin{ack}
We thank the referee for constructive comments that improved the clarity of this paper.
This work is supported in part by JSPS KAKENHI Grant Number 16K05296 (Y.T.), 
16H01101, and 16H03958 (T.N.).
K.I acknowledges support by the Spanish MINECO under grant AYA2016-76012-C3-1-P and
MDM-2014-0369 of ICCUB (Unidad de Excelencia `Mar\'ia de Maeztu').
This research is based on observations 
obtained with {\it XMM-Newton}, an ESA science mission 
with instruments and contributions directly funded by ESA Member States and NASA.
This research has made use of data obtained from the 3XMM XMM-Newton serendipitous source catalogue 
compiled by the 10 institutes of the XMM-Newton Survey Science Centre selected by ESA.
This work is based in part on observations made with the {\it Spitzer} Space Telescope, 
which is operated by the Jet Propulsion Laboratory, California Institute of Technology under a contract with NASA.
This research has made use of the NASA/ IPAC Infrared Science Archive, 
which is operated by the Jet Propulsion Laboratory, California Institute of Technology, under contract with NASA.
This work is based in part on data obtained as part of the UKIRT Infrared Deep Sky Survey.
This work is based in part on data collected at the Subaru Telescope and retrieved from the HSC data archive system, which is operated by the Subaru Telescope and Astronomy Data Center at National Astronomical Observatory of Japan.
 
The Hyper Suprime-Cam (HSC) collaboration includes the astronomical communities of Japan and Taiwan, and Princeton University.  
The HSC instrumentation and software were developed by the National Astronomical Observatory of Japan (NAOJ), the Kavli Institute for the Physics and Mathematics of the Universe (Kavli IPMU), the University of Tokyo, the High Energy Accelerator Research Organization (KEK), the Academia Sinica Institute for Astronomy and Astrophysics in Taiwan (ASIAA), and Princeton University.  Funding was contributed by the FIRST program from Japanese Cabinet Office, the Ministry of Education, Culture, Sports, Science and Technology (MEXT), the Japan Society for the Promotion of Science (JSPS),  Japan Science and Technology Agency  (JST),  the Toray Science  Foundation, NAOJ, Kavli IPMU, KEK, ASIAA,  and Princeton University.

The Pan-STARRS1 Surveys (PS1) have been made possible through contributions of the Institute for Astronomy, the University of Hawaii, the Pan-STARRS Project Office, the Max-Planck Society and its participating institutes, the Max Planck Institute for Astronomy, Heidelberg and the Max Planck Institute for Extraterrestrial Physics, Garching, The Johns Hopkins University, Durham University, the University of Edinburgh, Queen's University Belfast, the Harvard-Smithsonian Center for Astrophysics, the Las Cumbres Observatory Global Telescope Network Incorporated, the National Central University of Taiwan, the Space Telescope Science Institute, the National Aeronautics and Space Administration under Grant No. NNX08AR22G issued through the Planetary Science Division of the NASA Science Mission Directorate, the National Science Foundation under Grant No. AST-1238877, the University of Maryland, and Eotvos Lorand University (ELTE).
 
This paper makes use of software developed for the Large Synoptic Survey Telescope. We thank the LSST Project for making their code available as free software at http://dm.lsst.org.

\end{ack}

\appendix

\section*{$i$ band and 3.6 $\mu$m images}

$i$-band and 3.6 $\mu$m images of 49 objects, which are not shown in figure 2, are presented in the electronic version of this paper.

\begin{longtable}{llll}
\caption{The Data.}\label{table:sed}
\hline
Observatory & Instrument & Survey & Band  \\ 
\endhead
\hline
{\it XMM-Newton} & EPIC-PN, MOS & 3XMM-DR6 & 0.2$-$12 keV  \\
{\it Subaru}  & Hyper Supirme-Cam & SSP-Wide S15B & $g, r, i, z, y$\\
UKIRT & WFCAM  & DXS DR10plus& $J$, $K$ \\
	& WFCAM	& UDS DR10plus  & $J, H, K$ \\
{\it Spitzer} & IRAC & SWIRE & 3.6, 4.5, 5.8, 8.0 $\mu$m\\
		& MIPS & SWIRE & 24 $\mu$m\\
\hline
\end{longtable}

\begin{longtable}{cccccccc}
\caption{Results of SED fits.}\label{table:sed}
\hline
3XMM name & $z_{\rm phot}$ & $A_V$ & SED\footnotemark[$*$] & Class\footnotemark[$\dagger$] & $z_{\rm spec}$\footnotemark[$\ddagger$] & class\footnotemark[$\S$]  & Note\footnotemark[$\#$] \\
\hline
\endhead
\hline
\multicolumn{3}{l}{\hbox to 0pt{\parbox{180mm}{\footnotesize
\par \noindent
\footnotemark[$*$] Adopted SED template. ``+IR'' denotes the existence of infrared excess compared to model in fit without {\it Spitzer} data.
\par \noindent
\footnotemark[$\dagger$] Classification of SED type.
\par \noindent
\footnotemark[$\ddagger$] Spectroscopic redshift presented in Akiyama et al. (2015).
\par \noindent
\footnotemark[$\S$] Spectroscopic classification in Akiyama et al. (2015). NLA: Narrow-line AGN, BLA: Broad-line AGN.
\par \noindent
\footnotemark[$\#$] Note.
a: Redshift range in fit was ristricted to $1<z<2$.
b: {\it Spitzer} data was excluded in fit.\\
~~~c: Redshift range in fit was ristricted to $2<z<3$.
}}}
\endlastfoot
\hline
 J021502.3-034111 & 1.957 & 0.00 & Sd & SF & ... & ... & ... \\ %
 J021529.4-034233 & 1.262 & 0.50 & I20551 & AGN2 & ... & ... & a \\ %
 J021532.3-035124 & 2.099 & 1.10 & QSO1 & AGN1 & ... & ... & ... \\ %
 J021541.0-034505 & 1.460 & 0.10 & M82 & SF & ... & ... & ... \\ %
 J021606.6-050303 & 1.392 & 1.10 & I20551 & AGN2 & 0.471 & NLA & ... \\ %
 J021614.5-050351 & 2.048 & 0.80 & Mrk231 & AGN2 & 1.651 & BLA & ... \\ %
 J021625.7-050518 & 1.746 & 0.60 & I20551 & AGN2 & 1.873 & NLA & ... \\ %
 J021634.3-050724 & 1.136 & 0.50 & Sey2 & AGN2 & ... & ... & a, b \\ %
 J021642.3-043552 & 1.722 & 1.00 & Mrk231 & AGN2 & ... & ... & ... \\ %
 J021644.6-040651 & 1.085 & 0.10 & Mrk231 & AGN2 & ... & ... & a\\ %
 J021705.4-045655 & 1.112 & 1.00 & Mrk231 & AGN2 & ... & ... & ... \\ %
 J021705.7-052546 & 1.457 & 0.10 & Sdm+IR & SF & ... & ... & b \\ %
 J021721.2-052336 & 2.485 & 1.30 & BQSO1 & AGN1 & 1.382 & BLA & ... \\ %
 J021721.9-043655 & 1.035 & 1.20 & Torus & AGN2 & ... & ... & a\\ %
 J021725.8-051955 & 1.059 & 2.00 & QSO2 & AGN2 & ... & ... & ... \\ %
 J021729.3-052122 & 2.037 & 0.10 & I19254 & AGN2 & ... & ... & ... \\ %
 J021736.4-050106 & 1.169 & 0.00 & N6090+IR & SF & 1.423 & BLA & b \\ %
 J021744.1-034531 & 1.112 & 1.20 & Mrk231 & AGN2 & ... & ... & \\ %
 J021810.1-051844 & 2.066 & 0.70 & Sey18 & AGN2 & 2.523 & NLA & ... \\ %
 J021813.2-045051 & 0.649 & 1.80 & QSO2 & AGN2 & ... & ... & ... \\ %
 J021825.6-045945 & 1.478 & 0.40 & Mrk231 & AGN2 & ... & ... & ... \\ %
 J021842.8-051934 & 2.119 & 0.80 & Sdm+IR & SF & ... & ... & b\\ %
 J021914.8-045139 & 1.721 & 0.70 & I20551 & AGN2 & 1.626 & BLA & ... \\ %
 J022015.5-045654 & 1.957 & 1.20 & Sdm & SF & ... & ... & ... \\ %
 J022129.0-035359 & 1.366 & 1.20 & I20551 & AGN2 & ... & ... & ... \\ %
 J022145.5-034346 & 2.464 & 0.40 & I20551 & AGN2 & ... & ... & ... \\ %
 J022154.7-032558 & 2.420 & 0.50 & Mrk231 & AGN2 & ... & ... & ... \\ %
 J022205.0-033238 & 1.478 & 0.10 & Mrk231 & AGN2 & ... & ... & a \\ %
 J022214.4-034619 & 1.114 & 0.70 & I19254 & AGN2 & ... & ... & ... \\ %
 J022231.7-044910 & 2.081 & 1.40 & TQSO1 & AGN1 & ... & ... & ... \\ %
 J022314.5-041017 & 1.751 & 0.10 & Sey2 & AGN2 & ... & ... & ... \\ %
 J022326.5-041837 & 0.650 & 1.70 & QSO2 & AGN2 & ... & ... & ... \\ %
 J022330.8-044632 & 1.690 & 0.10 & Mrk231 & AGN2 & ... & ... & \\ %
 J022331.0-044234 & 2.406 & 1.10 & Mrk231 & AGN2 & ... & ... & c\\ %
 J022334.3-040841 & 1.262 & 1.40 & I20551 & AGN2 & ... & ... & ... \\ %
 J022337.9-040512 & 2.454 & 0.20 & Sd+IR & SF & ... & ... & b \\ %
 J022343.3-041622 & 1.320 & 0.90 & Mrk231 & AGN2 & ... & ... & a \\ %
 J022347.1-040051 & 2.710 & 1.50 & BQSO1 & AGN1 & ... & ... & ... \\ %
 J022352.0-052421 & 1.519 & 0.50 & Arp220 & SF & ... & ... & ... \\ %
 J022353.2-041532 & 1.509 & 0.20 & Arp220 & SF & ... & ... & ... \\ %
 J022358.2-050946 & 1.690 & 0.60 & I20551 & AGN2 & ... & ... & a\\ %
 J022405.2-041612 & 1.684 & 0.10 & Mrk231 & AGN2 & ... & ... & ... \\ %
 J022408.6-041151 & 1.668 & 0.90 & I22491 & AGN2 & ... & ... & ... \\ %
 J022410.3-040224 & 2.095 & 2.30 & TQSO1 & AGN1 & ... & ... & b\\ %
 J022412.5-035740 & 1.784 & 0.70 & I20551 & AGN2 & ... & ... & ... \\ %
 J022417.4-041812 & 1.644 & 1.20 & I22491+IR & AGN2 & ... & ... & b\\ %
 J022420.7-041224 & 1.385 & 1.10 & Sey18+IR & AGN2 & ... & ... & b\\ %
 J022421.1-040355 & 1.462 & 0.00 & M82 & SF & ... & ... & ... \\ %
 J022500.1-050831 & 1.020 & 0.40 & I19254 & AGN2 & ... & ... & ... \\ %
 J022504.5-043706 & 1.130 & 1.80 & QSO2 & AGN2 & ... & ... & ... \\ %
 J022510.6-043549 & 2.030 & 0.80 & I20551+IR & AGN2 & ... & ... & b \\ %
 J022624.3-041344 & 1.464 & 0.00 & Arp220+IR & SF & ... & ... & b \\ %
 J022625.2-044648 & 1.734 & 0.10 & Mrk231 & AGN2 & ... & ... & ... \\ %
\hline
\end{longtable}

\begin{table}
\begin{center}
\caption{Log of  {\it XMM-Newton} Observations}.
\label{table:obslog}
\begin{tabular}{clccc}
\hline 
Observation ID & Pointing\footnotemark[$*$] & \multicolumn{3}{c}{Exposure (ksec)}\footnotemark[$\dagger$]\\
	& 	& EPIC-MOS1 & EPIC-MOS2 & EPIC-PN \\
\hline
0037980601 & LSS6 & 13.2 & 13.3 & 7.5\\
0109520101 & XMDSOM1 & 25.8 & 25.8 & 19.5\\
0109520301 & XMDSOM3 & 21.8 & 21.8 & 16.0\\
0109520501 & XMDSOM5 & 24.0 & 24.0 & 17.7\\
0111110301 & XMDSSSC3 & 23.1 & 23.2 & 17.1\\
0111110401 & XMDSSSC4 & 28.0 & 28.1 & 17.4\\
0112370101 & SDS1 & 44.0 & 43.8 & 31.4\\
0112370301 & SDS2 & 43.0 & 42.6 & 23.5\\
0112370601 & SDS5 & 34.8 & 35.4 & 29.2\\
0112370701 & SDS6 & 48.7 & 48.7 & 37.3\\
0112370801 & SDS7 & 38.7 & 38.2 & 32.3\\
0112371001 & SDS1 & 46.0 & 45.6 & 33.5\\
0112372001 & SDS4 & 27.3 & 27.3 & 23.3\\
0112680201 & MLS2 & 11.9 & 11.9 & 7.5\\
0112681001 & MLS7 & 24.0 & 24.0 & 17.2\\
0210490101 & XLSSJ022404.0$-$04132 & 83.9 & 82.9 & 67.4\\
0404967901 & LSS65 & 14.4 & 14.4 & 11.4\\
0553910601 & LSS37 & 14.3 & 14.3 & 12.3\\
0553911601 & LSS61 & 13.0 & 13.0 & 10.1\\
0604280101 & XLSSC006 & 97.4 & 97.1 & 82.5\\
0673110201 & XMM-129 & 32.3 & 32.3 & 23.6\\
\hline
\multicolumn{5}{@{}l@{}}{\hbox to 0pt{\parbox{85mm}{\footnotesize
Notes.
\par\indent
\footnotemark[$*$] Targets in the original proposal.
\par\indent
\footnotemark[$\dagger$] Net exposure time after data screening.
}\hss}}
\end{tabular}
\end{center}
\end{table}

\begin{longtable}{cccccccc}
\caption{Results of power law fits to X-ray spectra}\label{table:pl}
\label{table:pl}
\hline
3XMM Name & Observation ID & Net counts\footnotemark[$*$] &  $N_{\rm H}$\footnotemark[$\dagger$] & Photon index & Flux\footnotemark[$\ddagger$] & $C$/dof\footnotemark[$\S$] & Note\footnotemark[$\|$]\\
	&		&	& ($10^{20}$ cm$^{-2}$) &  & ($10^{-14}$ erg s$^{-1}$ cm$^{-2}$)\\
\hline
\endhead
\hline
\endfoot
\hline
\multicolumn{8}{@{}l@{}}{\hbox to 0pt{\parbox{180mm}{\footnotesize
Notes.
\par\indent
\footnotemark[$*$] Net counts in the 0.5--10 keV band. Counts from all the available sensors are summed.
\par\indent
\footnotemark[$\dagger$] Galactic absorption column density in units of $10^{20}$ cm$^{-2}$.
\par\noindent
\footnotemark[$\ddagger$] Observed flux in the 2--10 keV band in units of $10^{-14}$ erg s$^{-1}$.
\par\noindent
\footnotemark[$\S$] $C$ statistic/degrees of freedom.
\par\noindent
\footnotemark[$\|$]
a: Different normalization of power law for EPIC-PN and EPIC-MOS are allowd in fit. Flux is for EPIC-PN.
b: Only EPIC-PN is used. At chip boundary in EPIC-MOS1 and EPIC-MOS2.
c: Only EPIC-PN and EPIC-MOS1 are used.Out of field of view of EPIC-MOS2.
d: Only EPIC-PN and EPIC-MOS1 are used. Located at chip boundary in EPIC-MOS1.
e: Only EPIC-PN is used. Not clearly visible in EPIC-MOS1 image and located at chip boundary in EPIC-MOS2. 
f: Only EPIC-PN is used. Out of field of view of EPIC-MOS1 and at edge of field of view of EPIC-MOS2.
}\hss}}
\endlastfoot
J021502.3-034111 & 0673110201	& 66 & 2.07	& $1.14^{+0.39}_{-0.36}$ & 1.16 & 110.8/113 & a\\
J021529.4-034233 & 0673110201	& 230 & 2.07	& $1.74^{+0.17}_{-0.17}$ & 1.69 & 296.5/281 & \\
J021532.3-035124 & 0673110201	& 72 & 2.04	& $1.87^{+0.39}_{-0.36}$ & 0.81 & 116.4/118 & \\
J021541.0-034505 & 0673110201	& 87 & 2.07	& $1.94^{+0.34}_{-0.31}$ & 0.54 & 142.3/139 & \\
J021606.6-050303 & 0112370601	& 89 & 2.06	& $1.41^{+0.30}_{-0.29}$ & 1.18 & 139.3/125 & \\
J021614.5-050351 & 0112370601	& 88 & 2.06	& $1.52^{+0.31}_{-0.39}$ & 0.94 & 121.1/122 & \\
J021625.7-050518 & 0112370601	& 93 & 2.05	& $1.77^{+0.27}_{-0.25}$ & 1.18 & 100.7/122 & b\\
J021634.3-050724 & 0112370601	& 148 & 2.05	& $1.54^{+0.22}_{-0.21}$ & 1.90 & 190.3/217 & \\
J021642.3-043552 & 0112372001	& 659 & 1.97	& $1.11^{+0.09}_{-0.09}$ & 17.10 & 583.6/531 & \\
J021644.6-040651 & 0553911601	& 144 & 1.99	& $1.46^{+0.20}_{-0.20}$ & 5.81 & 203.0/151 & \\
J021705.4-045655 & 0112371001	& 78  & 2.05	& $1.43^{+0.34}_{-0.32}$ & 3.00 & 108.8/107 & c\\
J021705.7-052546 & 0112370701	& 152 & 2.23	& $1.27^{+0.19}_{-0.19}$ & 1.68 & 149.8/180 & \\
J021721.2-052336 & 0112370701	& 77 & 2.22	& $1.88^{+0.36}_{-0.33}$ & 0.41 & 127.6/129 & \\
J021721.9-043655 & 0112372001	& 133 & 1.97	& $1.40^{+0.22}_{-0.21}$ & 1.77 & 147.7/161 & \\
J021725.8-051955 & 0112370701	& 97 &  2.22	& $1.51^{+0.33}_{-0.31}$ & 0.60 & 170.0/145 & \\
J021729.3-052122 & 0112370701	& 96 & 2.22	& $0.90^{+0.27}_{-0.27}$ & 1.31 & 146.0/137 & \\
J021736.4-050106 & 0112371001	& 111 & 1.98	& $1.36^{+0.26}_{-0.25}$ & 1.22 & 163.4/157 & \\
J021744.1-034531 & 0404967901	& 43 & 2.05	& $2.03^{+0.43}_{-0.40}$ & 0.84 & 81.6/83 & \\
J021810.1-051844 & 0112370801	& 93 & 2.22	& $1.63^{+0.32}_{-0.30}$ & 1.23 & 109.3/137 & \\
J021813.2-045051 & 0112371001	& 106 & 1.99	& $1.11^{+0.27}_{-0.25}$ & 2.32 & 161.5/137 & d\\
J021825.6-045945 & 0112370101	& 155 & 1.99	& $0.71^{+0.21}_{-0.21}$ & 3.34 & 204.7/203 & \\
J021842.8-051934 & 0112370801	& 205 & 2.13	& $0.95^{+0.17}_{-0.17}$ & 2.86 & 206.9/246 & \\
J021914.8-045139 & 0112370301	& 116 & 2.01	& $1.64^{+0.25}_{-0.24}$ & 1.66 & 147.9/159 & \\
J022015.5-045654 & 0112370301	& 148 & 2.02	& $1.31^{+0.22}_{-0.22}$ & 2.86 & 183.4/177 & \\
J022129.0-035359 & 0604280101	& 100 & 2.16	& $1.17^{+0.57}_{-0.52}$ & 1.40 & 480.2/556 & \\
J022145.5-034346 & 0604280101	& 273 & 2.16	& $1.14^{+0.22}_{-0.21}$ & 1.35 & 503.4/646 & \\
J022154.7-032558 & 0037980601	& 108 & 2.21	& $1.55^{+0.24}_{-0.23}$ & 3.39 & 157.8/138 & \\
J022205.0-033238 & 0604280101	& 131 & 2.22	& $0.68^{+0.32}_{-0.33}$ & 2.78 & 504.4/564 & \\
J022214.4-034619 & 0604280101	& 410 & 2.16	& $1.66^{+0.18}_{-0.17}$ & 1.33 & 663.9/801 & \\
J022231.7-044910 & 0109520501	& 36 & 2.10	& $1.29^{+0.40}_{-0.38}$ & 3.44 & 41.4/45 & e\\
J022314.5-041017 & 0109520101	& 52 & 2.22	& $1.67^{+0.40}_{-0.37}$ & 0.53 & 76.6/81 & \\
J022326.5-041837 & 0210490101	& 102 & 2.22	& $1.40^{+0.34}_{-0.32}$ & 0.93 & 153.11/197 & \\
J022330.8-044632 & 0109520501	& 363 & 2.18	& $1.40^{+0.12}_{-0.12}$ & 5.81 & 313.5/328 & \\
J022331.0-044234 & 0109520501	& 79 & 2.18	& $1.41^{+0.28}_{-0.27}$ & 2.02 & 83.0/95 & \\
J022334.3-040841 & 0210490101	& 88 & 2.27	& $0.35^{+0.36}_{-0.39}$ & 1.41 & 184.8/174 & \\
J022337.9-040512 & 0210490101	& 212 & 2.27	& $1.25^{+0.19}_{-0.18}$ & 1.84 & 239.6/260 & \\
J022343.3-041622 & 0210490101	& 108 & 2.22	& $1.35^{+0.28}_{-0.27}$ & 0.82 & 169.8/181 & \\
J022347.1-040051 & 0210490101	& 107 & 2.29	& $1.36^{+0.33}_{-0.31}$ & 1.53 & 172.9/189 & \\
J022352.0-052421 & 0553910601	& 75 & 2.22	& $1.91^{+0.33}_{-0.31}$ & 2.29 & 99.1/98 & \\
J022353.2-041532 & 0210490101	& 152 & 2.22	& $1.63^{+0.23}_{-0.22}$ & 0.64 & 249.4/235 & \\
J022358.2-050946 & 0111110401	& 77 & 2.23	& $1.72^{+0.31}_{-0.29}$ & 0.78 & 85.7/90 & \\
J022405.2-041612 & 0210490101	& 387 & 2.27	& $1.80^{+0.14}_{-0.13}$ & 1.06 & 338.9/440 & \\
J022408.6-041151 & 0210490101	& 142 & 2.27	& $0.40^{+0.25}_{-0.26}$ & 1.52 & 237.6/253 & \\
J022410.3-040224 & 0210490101	& 68 & 2.27	& $1.75^{+0.37}_{-0.34}$ & 0.53 & 131.3/118 & \\
J022412.5-035740 & 0210490101	& 35 & 2.29	& $1.56^{+0.80}_{-0.62}$ & 0.86 & 95.7/72 & f\\
J022417.4-041812 & 0210490101	& 57 & 2.27	& $1.21^{+0.43}_{-0.41}$ & 0.51 & 129.7/144 & \\
J022420.7-041224 & 0210490101	& 223 & 2.27	& $1.91^{+0.19}_{-0.18}$ & 0.58 & 247.0/302 & \\
J022421.1-040355 & 0210490101	& 168 & 2.27	& $1.37^{+0.22}_{-0.21}$ & 1.45 & 252.9/259 & \\
J022500.1-050831 & 0111110301	& 96 & 2.27	& $1.21^{+0.28}_{-0.27}$ & 0.78 & 125.4/114 & \\
J022504.5-043706 & 0112681001	& 69 & 2.31	& $0.96^{+0.30}_{-0.30}$ & 3.08 & 76.4/91 & \\
J022510.6-043549 & 0112681001	& 240 & 2.31	& $1.48^{+0.16}_{-0.16}$ & 3.65 & 244.9/260 & \\
J022624.3-041344 & 0112680201	& 64 & 2.34	& $2.13^{+0.39}_{-0.35}$ & 1.21 & 79.7/89 & \\
J022625.2-044648 & 0109520301	& 130 & 2.29	& $1.31^{+0.20}_{-0.20}$ & 3.75 & 114.6/124 & \\
\hline
\end{longtable}

\begin{longtable}{ccccc}
\caption{Results of absorbed power law fits to X-ray spectra.}
\label{table:abs}
\hline
3XMM Name\footnotemark[$*$] & $N_{\rm H}$\footnotemark[$\dagger$] & $L_{2-10}$
\footnotemark[$\ddagger$] & $C$/dof\footnotemark[$\S$] & \\
	& ($10^{22}$ cm$^{-2}$) & ($10^{44}$ erg s$^{-1}$) & \\
\hline
\endhead
\hline
\endfoot
\hline
\multicolumn{5}{@{}l@{}}{\hbox to 0pt{\parbox{85mm}{\footnotesize
Notes.
\par\indent
\footnotemark[$*$] Observation ID and assumed Galactic absorption column density are same as in table \ref{table:pl}
\par\indent
\footnotemark[$\dagger$] Intrinsic absorption column density at source redshift in units of $10^{22}$ cm$^{-2}$. 
Photometric or spectroscopic (when available) redshifts are assumed in the spectral fits.
\par\indent
\footnotemark[$\ddagger$] X-ray luminosity in 2--10 keV corrected for the Galactic and intrinsic absorption in units of $10^{44}$ erg s$^{-1}$. 
Photometric or spectroscopic (when available) redshifts are assumed.
\par\indent
\footnotemark[$\S$] $C$ statistic/degrees of freedom.
}\hss}}
\endlastfoot
J021502.3-034111 & $5.8^{+4.3}_{-3.1}$ & 2.30& 101.7/113 & \\
J021529.4-034233 & $0.069^{+0.15}_{-0.069}$ & 1.21 & 298.1/281 & \\
J021532.3-035124 & $0.17^{+0.73}_{-0.17}$ & 2.38 & 116.2/118 & \\
J021541.0-034505 & $0.051^{+0.34}_{-0.051}$ & 0.721 & 142.3/139 & \\
J021606.6-050303 & $0.17^{+0.18}_{-0.13}$ & 0.0613 & 140.7/125 & \\
J021614.5-050351 & $0.84^{+0.82}_{-0.56}$ & 1.19 & 117.6/122 & \\
J021625.7-050518 & $0.17^{+0.41}_{-0.17}$ & 2.29 & 100.7/122 & \\
J021634.3-050724 & $0.23^{+0.22}_{-0.18}$ & 0.892 & 192.8/215 & \\
J021642.3-043552 & $3.3^{+0.53}_{-0.47}$ & 21.6 & 465.6/531 & \\
J021644.6-040651 & $0.94^{+0.5}_{-0.55}$ & 2.10 & 188.1/151 & \\
J021705.4-045655 & $0.39^{+0.42}_{-0.29}$ & 1.29 & 108.5/107 & \\
J021705.7-052546 & $1.1^{+0.54}_{-0.42}$ & 1.27 & 150.2/180 & \\
J021721.2-052336 & $0^{+0.14}_{-0}$ & 0.423 & 127.6/129 & \\
J021721.9-043655 & $0.92^{+0.55}_{-0.40}$ & 0.744 & 137.0/161 & \\
J021725.8-051955 & $0.40^{+0.52}_{-0.37}$ & 0.247 & 170.5/145 & \\
J021729.3-052122 & $6.5^{+3.5}_{-2.5}$ & 2.17 & 138.8/137 & \\
J021736.4-050106 & $1.00^{+0.81}_{-0.58}$ & 1.00 & 161.9/157 & \\
J021744.1-034531 & $0^{+0.083}_{-0}$ & 0.65 & 81.9/83 & \\
J021810.1-051844 & $0.7^{+1.2}_{-0.7}$ & 4.3 & 109.6/137 & \\
J021813.2-045051 & $0.81^{+0.43}_{-0.31}$ & 0.261 & 147.3/137 & \\
J021825.6-045945 & $9.5^{+3}_{-2.4}$ & 3.11 & 196.5/203 & \\
J021842.8-051934 & $7.3^{+2.8}_{-2.2}$ & 5.49 & 200.7/246 & \\
J021914.8-045139 & $0.19^{+0.34}_{-0.19}$ & 1.96 & 149.6/159 & \\
J022015.5-045654 & $2.2^{+1.3}_{-0.94}$ & 4.80 & 177.4/177 & \\
J022129.0-035359 & $0.65^{+1.6}_{-0.62}$ & 0.659 & 481.4/556 & \\
J022145.5-034346 & $3.6^{+2.0}_{-1.4}$ & 3.24 & 495.5/646 & \\
J022154.7-032558 & $0.84^{+0.98}_{-0.72}$ & 9.75 & 159.4/137 & \\
J022205.0-033238 & $7.2^{+3.7}_{-4.5}$ & 2.21 & 491.7/564 & \\
J022214.4-034619 & $0.2^{+0.14}_{-0.12}$ & 0.688 & 660.0/801 & \\
J022231.7-044910 & $2.2^{+2.4}_{-1.5}$ & 6.63 & 39.0/45 & \\
J022314.5-041017 & $0.17^{+0.67}_{-0.17}$ & 0.757 & 77.3/81 & \\
J022326.5-041837 & $0.064^{+0.21}_{-0.064}$ & 0.0848 & 158.4/197 & \\
J022330.8-044632 & $1.4^{+0.40}_{-0.35}$ & 7.65 & 280.2/328 & \\
J022331.0-044234 & $2.0^{+1.6}_{-1.1}$ & 5.80 & 79.4/95 & \\
J022334.3-040841 & $12^{+7.6}_{-5.1}$ & 0.861 & 171.3/174 & \\
J022337.9-040512 & $3.2^{+1.5}_{-1.1}$ & 5.00 & 233.0/260 & \\
J022343.3-041622 & $0.51^{+0.48}_{-0.11}$ & 0.437 & 174.5/181 & \\
J022347.1-040051 & $1.5^{+2.2}_{-1.3}$ & 4.70 & 176.4/189 & \\
J022352.0-052421 & $0.12^{+0.38}_{-0.12}$ & 3.28 & 98.7/98 & \\
J022353.2-041532 & $0.54^{+0.51}_{-0.37}$ & 0.721 & 246.1/235 & \\
J022358.2-050946 & $0.21^{+0.65}_{-0.21}$ & 1.14 & 86.1/90 & \\
J022405.2-041612 & $0.16^{+0.19}_{-0.16}$ & 1.70 & 337.9/440 & \\
J022408.6-041151 & $18^{+10}_{-6}$ & 1.84 & 232.5/253 & \\
J022410.3-040224 & $0.39^{+1.4}_{-0.39}$ & 1.38 & 131.2/118 & \\
J022412.5-035740 & $0.46^{+1.2}_{-0.46}$ & 1.21 & 95.4/72 & \\
J022417.4-041812 & $3.8^{+5.5}_{-3.3}$ & 0.516 & 130.1/144 & \\
J022420.7-041224 & $0.15^{+0.20}_{-0.15}$ & 0.687 & 244.7/302 & \\
J022421.1-040355 & $1.1^{+0.68}_{-0.51}$ & 1.25 & 246.1/259 & \\
J022500.1-050831 & $1.5^{+0.89}_{-0.65}$ & 0.836 & 116.8/114 & \\
J022504.5-043706 & $2.2^{+1.5}_{-1.1}$ & 1.24 & 77.7/91 & \\
J022510.6-043549 & $1.4^{+0.61}_{-0.5}$ & 7.59 & 232.0/260 & \\
J022624.3-041344 & $0^{+0.23}_{-0}$ & 2.04 & 80.5/89 & \\
J022625.2-044648 & $1.1^{+0.64}_{-0.51}$ & 4.39 & 120.2/124 & \\
\hline
\end{longtable}

\begin{longtable}{ccccccc}
\caption{Flux ratios.}
\label{table:ratio}
\hline
3XMM Name & $i$ mag\footnotemark[$*$] & $F_{\rm X}/F_i$\footnotemark[$\dagger$] & $F_{\rm X}/\lambda F_{\lambda, 24}$ & 
$i-[3.6]$ & $[3.6]-[4.5]$ & $i-[24]$ \\
\hline
\endhead
\hline
\endfoot
\hline
\multicolumn{7}{@{}l@{}}{\hbox to 0pt{\parbox{180mm}{\footnotesize
Notes.
\par\indent
\footnotemark[$*$] cmodel magnitude in $i$ band. Galactic extinction is not corrected.
\par\noindent
\footnotemark[$\dagger$] X-ray to $i$-band flux ratio.
$F_{\rm X}$ is X-ray flux in the 2--10 keV band corrected for the Galactic absorption, 
and $F_i$ is $\Delta \lambda F_{\lambda}$ \\
at $i$ band corrected for the Galactic extinction.
\par\noindent
\footnotemark[$\ddagger$] X-ray to 24 $\mu$m flux ratio. 
$F_{\rm X}$ is X-ray flux in the 2--10 keV band
corrected for the Galactic absorption, and $F_{\lambda}$ is flux \\
density at 24$\mu$m in units of erg s$^{-1}$ cm$^{-2}$ \AA$^{-1}$.
}\hss}}
\endlastfoot
J021502.3-034111 & 24.086 & 13.8 & ...  &3.22 & 0.17 & ... \\
J021529.4-034233 & 23.565 & 13.5 & ...  &3.14 & 0.11 & ... \\
J021532.3-035124 & 23.613 & 8.2 & ...  &2.49 & 0.25 & ... \\
J021541.0-034505 & 23.553 & 5.6 & ...  &3.29 & 0.14 & ... \\
J021606.6-050303 & 24.218 & 13.3 & ...  &3.94 & 0.26 & ... \\
J021614.5-050351 & 25.240 & 32.4 & ...  &5.19 & 0.51 & ... \\
J021625.7-050518 & 24.114 & 16.4 & ...  &3.52 & 0.20 & ... \\
J021634.3-050724 & 24.744 & 37.9 & ...  &4.24 & 0.46 & ... \\
J021642.3-043552 & 23.951 & 158.6 & 7.19 & 4.90 & 0.39 & 7.80 \\
J021644.6-040651 & 23.596 & 34.3 & ...  &3.11 & 0.20 & ... \\
J021705.4-045655 & 24.497 & 44.2 & 2.61 & 4.57 & 0.47 & 7.51 \\
J021705.7-052546 & 24.369 & 20.8 & 1.28 & 3.92 & 0.68 & 7.46 \\
J021721.2-052336 & 23.670 & 4.2 & ...  &2.41 & 0.29 & ... \\
J021721.9-043655 & 26.202 & 151.2 & 0.87  & 7.20 & 0.68 & 10.04 \\
J021725.8-051955 & 25.244 & 19.6 & ...  &4.61 & 0.20 & ... \\
J021729.3-052122 & 26.010 & 69.6 & ...  &4.75 & 0.47 & ... \\
J021736.4-050106 & 23.573 & 8.0 & 1.22 & 3.85 & 0.30 & 6.48 \\
J021744.1-034531 & 24.715 & 28.6 & ...  &5.27 & 0.14 & ... \\
J021810.1-051844 & 24.434 & 20.4 & ...  &3.73 & 0.21 & ... \\
J021813.2-045051 & 24.398 & 30.7 & ...  &3.88 & 0.34 & ... \\
J021825.6-045945 & 24.162 & 35.4 & ...  &4.24 & 0.05 & ... \\
J021842.8-051934 & 25.331 & 86.1 & ...  &4.69 & 0.31 & ... \\
J021914.8-045139 & 23.799 & 14.9 & ...  &3.45 & 0.36 & ... \\
J022015.5-045654 & 24.627 & 49.1 & ...  &4.51 & 0.25 & ... \\
J022129.0-035359 & 24.335 & 12.2 & ...  &4.12 & 0.24 & ... \\
J022145.5-034346 & 24.367 & 15.0 & ...  &3.70 & 0.01 & ... \\
J022154.7-032558 & 24.380 & 48.4 & ...  &4.16 & 0.26 & ... \\
J022205.0-033238 & 24.237 & 29.0 & ...  &3.52 & 0.32 & ... \\
J022214.4-034619 & 25.119 & 43.4 & ...  &4.43 & 0.31 & ... \\
J022231.7-044910 & 23.821 & 27.8 & 2.71  & 3.37 & 0.52 & 6.97 \\
J022314.5-041017 & 25.063 & 15.2 & ...  &4.18 & 0.09 & ... \\
J022326.5-041837 & 23.814 & 5.9 & ...  &3.69 & 0.38 & ... \\
J022330.8-044632 & 23.878 & 55.5 & ...  &3.08 & 0.79 & ... \\
J022331.0-044234 & 26.351 & 176.7 & ...  &4.82 & 0.42 & ... \\
J022334.3-040841 & 24.467 & 18.8 & ...  &4.48 & 0.15 & ... \\
J022337.9-040512 & 24.523 & 26.9 & 0.56  & 4.81 & 0.69 & 8.65 \\
J022343.3-041622 & 25.247 & 21.8 & ...  &4.93 & 0.50 & ... \\
J022347.1-040051 & 23.749 & 9.9 & ...  &3.22 & 0.11 & ... \\
J022352.0-052421 & 24.394 & 50.3 & ...  &4.21 & 0.12 & ... \\
J022353.2-041532 & 24.147 & 8.8 & ...  &3.62 & 0.19 & ... \\
J022358.2-050946 & 24.381 & 13.3 & ...  &3.58 & 0.37 & ... \\
J022405.2-041612 & 24.118 & 15.6 & ...  &3.02 & 0.53 & ... \\
J022408.6-041151 & 23.836 & 11.4 & ...  &4.03 & 0.23 & ... \\
J022410.3-040224 & 25.303 & 22.4 & ...  &4.40 & 0.41 & ... \\
J022412.5-035740 & 23.629 & 6.2 & ...  &3.72 & 0.18 & ... \\
J022417.4-041812 & 23.952 & 4.2 & 0.57 & 4.31 & 0.35 & 6.60 \\
J022420.7-041224 & 24.165 & 10.6 & ...  &4.10 & 0.41 & ... \\
J022421.1-040355 & 24.304 & 19.0 & ...  &3.30 & 0.27 & ... \\
J022500.1-050831 & 24.983 & 55.9 & ...  &4.05 & 0.25 & ... \\
J022504.5-043706 & 24.539 & 43.0 & 3.14  & 5.03 & 0.34 & 7.28 \\
J022510.6-043549 & 24.816 & 84.9 & ...  &4.46 & 0.44 & ... \\
J022624.3-041344 & 24.348 & 32.6 & 2.07 & 4.04 & 0.70 & 7.43 \\
J022625.2-044648 & 23.791 & 27.7 & ...  &3.35 & 0.58 & ... \\
\hline
\end{longtable}

\end{document}